\documentclass{aa}  
\usepackage[colorlinks=true, linkcolor=blue, citecolor=blue, urlcolor=blue]{hyperref}

\usepackage{graphicx}
\usepackage{txfonts}

\newcommand{\app}[1]{Appendix~\ref{app:#1}}

\newcommand{\se}[1]{Section~\ref{sec:#1}}

\newcommand{\fg}[1]{Figure~\ref{fig:#1}}
\newcommand{\Fg}[1]{Figure~\ref{fig:#1}}

\newcommand{\Tb}[1]{Table~\ref{tab:#1}}
\newcommand{\eq}[1]{Equation~(\ref{eq:#1})}
\newcommand{\Eq}[1]{Equation~(\ref{eq:#1})}

\usepackage{xcolor,bm}
\usepackage{CJKutf8}

\newcommand{\rebound}{\texttt{Rebound}\xspace}
\newcommand{\hr}{HR 8799\xspace}
\newcommand{\pds}{PDS 70\xspace}
\newcommand{\gj}{GJ 876\xspace}
\newcommand{\Po}{Fixed point orbits\xspace}
\newcommand{\Fpos}{Fixed point orbits\xspace}
\newcommand{\fpo}{fPO\xspace}
\newcommand{\fpos}{fPOs\xspace}
\newcommand{\po}{fPO\xspace}
\newcommand{\pos}{fPOs\xspace}

\begin{document}

\title{Fitting Resonant Exoplanets Orbits. I.}
\subtitle{Properties and Applications of Fixed-Point Orbits}

\author{Chris W. Ormel\inst{1}\thanks{Corresponding author: \href{chrisormel@tsinghua.edu.cn}{chrisormel@tsinghua.edu.cn}}
	\and
	Tian Yi (\begin{CJK*}{UTF8}{gbsn}易天\end{CJK*}) \inst{1}
	\and
	Shuo Huang (\begin{CJK*}{UTF8}{gbsn}黄硕\end{CJK*}) \inst{1,2}
}
\institute{Department of Astronomy, Tsinghua University, Haidian DS 100084 Beijing, China
	\and
	Tsung-Dao Lee Institute, Shanghai Jiao-Tong University, 1 Lisuo Road, Shanghai 201210, China }
\date{\today}
\abstract{
	Intact chains of mean-motion resonances may preserve information about the formation and early dynamical evolution of multiplanet systems. Although most mature multiplanet systems are non-resonant, the resonant fraction may have been substantially higher at young ages. Establishing whether exoplanet data are consistent with a resonant configuration is often challenging because the planetary orbits must be sufficiently well constrained. Moreover, while resonance configurations promote stability, standard orbital fitting methods do not impose such a dynamical constraint, with stability tested only a posteriori. As the first step in a series of papers aimed at favoring stable orbital solutions during the fitting process, we develop a method that restricts solutions to stable fixed-point orbits (\fpos)---coplanar, exactly periodic configurations associated with resonant dynamics.
	We show that dissipative evolution drive resonant systems toward an \fpo state, motivating for \fpo-constrained fitting for those planet systems that have avoided large-scale dynamical rearrangements since formation.
	The \fpo model greatly reduces the number of free parameters and yields stable solutions by construction. We determine the best-fitting \fpos through $\chi^2$ minimization and apply the method to astrometric data for the directly imaged \pds and \hr systems and to radial-velocity data for \gj. For \hr we independently recover a previoulsy reported resonant solution, illustrating the fidelity of our method. For \pds we find solutions fall in two groups, with the group where the outer planet c mass exceeds planet b more aligned with previous measurements. Finally, For \gj our \fpo fit is on par with previous unconstrained fitting, which strengthens the case that its planets reside deep in resonance. }

\keywords{ 
    Planetary systems --
    Planets and satellites: dynamical evolution and stability -- 
    Planets and satellites: formation --
    Stars: individual: HR 8799, PDS 70, GJ 876
}

\maketitle

\section{Introduction}
Driven by extensive ground- and space-based surveys employing radial velocities, transits, astrometry, and microlensing, the number of confirmed exoplanets has surpassed 6,000 and is expected to climb substantially in forthcoming data releases and missions, including Gaia DR4 and the Roman, PLATO, and Earth 2.0 (ET) \citep{GaiaCollaborationEtal2016,PennyEtal2019,GeEtal2022,RauerEtal2025}. The number of known multiplanet systems has likewise grown dramatically. Among these, systems containing planets in mean-motion resonance (MMR) are particularly informative, as they retain imprints of the physical processes that operated during planet formation. Specifically, convergent orbital migration ensures trapping in first-order MMRs given sufficient eccentricity damping, both of which arise naturally from interactions with a gas-rich protoplanetary disk.  Conversely, the present-day dynamical state of a resonant system can be used to infer the conditions that prevailed at early times, as has been done, for example, for TRAPPIST-1 \citep{HuangOrmel2022,PichierriEtal2024}.

Surveys such as Kepler have shown that the majority of adjacent planets pairs are neither in nor near low-order MMRs \citep{LissauerEtal2011i,WinnFabrycky2015}. Nevertheless, many planet systems---perhaps even the majority---could have emerged from the gas-disk phase as resonance chains. In the "breaking-the-chains" model resonance chains are a temporal phase, with many systems becoming unstable after disk dispersal \citep{IzidoroEtal2017,IzidoroEtal2021}. Recent analyses also provides observational support for this picture: the fraction of planet pairs in (or close to) low-order commensurability was much higher among systems that are $10{-}100$ Myr old \citep{HamerSchlaufman2024,DaiEtal2024}. Solidifying this temporal dependence and determining whether these young systems are genuinely resonant is important for understanding the processes by which planets enter and escape from resonance \citep{HuEtal2025}.

But how is it determined whether planets are in resonance? The crudest criterion is the proximity of the planets' period ratio to an integer commensurability, such as $j{+}1{:}j$ for a first-order resonance. However, proximity alone does not establish resonance from a dynamical perspective. More commonly, resonance is associated with libration of an appropriate angle, e.g.,
\begin{equation} \label{eq:phi-classical}
	\phi = (j+1)\lambda_2 -j\lambda_1 -\varpi_\mathrm{eff}
\end{equation}
for first-order resonances, where $\lambda_i$ is the mean longitude of planet $i$ and $\varpi_\mathrm{eff}$ is an effective longitude of periapsis.\footnote{For the classical resonant angles ($\phi=\phi_i$) $\varpi_\mathrm{eff}=\varpi_i$ is just the longitude of periapsis of one of the planets. More formally $\varpi_\mathrm{eff}$ for a two-planet system is the argument of a weighted linear combination of the planets' complex eccentricities \citep[e.g.][]{PetitEtal2020i}. A stricter definition further requires the trajectory to lie within a resonant domain bounded by a separatrix \citep{HenrardLemaitre1983,BatyginMorbidelli2013}. In this Paper we do not adopt this stricter distinction and instead equate libration of the relevant resonant angle with resonance.} Therefore, establishing the resonant nature of a system requires us to characterize the planets' orbits, which can be challenging when only a limited number of observations are available.

Orbit fitting is commonly performed using Bayesian sampling packages such as \texttt{orbitize!} \citep{BluntEtal2020}, \texttt{orvara} \citep{BrandtEtal2021}, or \texttt{octofitter} \citep{ThompsonEtal2023i}. For a multiplanet systems this involves many unknowns---at least six orbital elements and a mass for each planet---making a computationally efficient evaluation of the likelihood function essential. One concern is that the posterior sample may contain many solutions that are dynamically unstable on timescales longer than the time span of the observation but shorter than the age of the system. The scarcity of long-term stable orbits generated by these MCMC searches is particularly acute for directly imaged systems: the astrometric observations cover only small arcs of the planets' orbits, allowing for a large number of good fits, while the large planet masses in these systems render much of this phase space unstable. Naturally, the problem compounds as the number of planets increases \citep[e.g.][]{FabryckyMurray-Clay2010,WangEtal2018}. A second, subtler concern is that searches based on (log-)uniform sampling of orbital elemens would undersample resonant solutions. This is because such configurations occupy only a small fraction of the phase-space volume. These two concerns are related: in multiplanet systems, resonance can protect configurations that would otherwise be unstable.

In this paper series, we will pursue a strategy of preselecting resonant configurations and thereby favoring stability when fitting orbital data. One possibility is to construct a prior by numerically integrating each trial solution for several hundred to $10^4$ orbital periods. For example, \citet{JensenMillholland2022} considered priors favoring small libration amplitudes of the Laplace resonance angle to mitigate the phase-space-volume bias. The MEGNO chaos indicator \citep{GozdziewskiEtal2005} or the \texttt{SPOCK} stability classifier \citep{TamayoEtal2020} could likewise be evaluated for each trial solution. Although substantially faster than integrations over the lifetime of the system, these methods can remain computationally prohibitive for complex MCMC searches.
We defer such a Bayesian treatment to a follow-up paper, in which we will develop an analytic ``resonance prior''. In this work, we take a cleaner, more restrictive first step: we constrain the orbital solutions to fixed-point orbits (fPOs) and determine the best-fitting configurations through $\chi^2$ minimization.

An \fpo is a resonant periodic solution whose planetary positions and velocities repeat after a common period ($t_\mathrm{po}$), up to an overall rotation of the system.\footnote{In the celestial mechanics literature, these solutions are conventionally termed ``periodic orbits'' \citep[e.g.][]{Hadjidemetriou2006,AntoniadouVoyatzis2014}. Outside this specific usage, however, orbital motion can exhibit periodicity on several distinct timescales, including the orbital and libration periods. We prefer the term ``fixed-point orbit'' to emphasize the immutable nature of the orbit.} The precession rate and $t_\mathrm{po}$ of the \fpo completely specify its future orbit state.  The stability of an \fpo can be readily assessed by perturbing the solution and determining whether the departure grows with time. Fixed-point orbits arise naturally as outcomes of dissipative evolution of resonant systems \citep{GoldreichSchlichting2014}. During planet formation, eccentricity damping due to planet-disk interactions can reduce resonance librations and drive a system to a fixed point. After disk dispersal, planet-star tidal interactions can play a similar role \citep{LithwickWu2012}. Stable \fpos therefore provide physically motivated templates for fitting planetary orbits.

Models based on fixed-point orbits have previously been used to fit the \hr system \citep{GozdziewskiMigaszewski2020,ZurloEtal2022}. Here, we further develop the \po approach and apply it to additional resonances, planet systems (\pds and \gj), and other types of observational data, including radial velocities (RVs). 
Our goal is to demonstrate the utility of \fpos for fitting resonant orbital configurations to exoplanet data, thereby identifying regions of parameter space where the fitted orbits are guaranteed to be dynamically stable. 
This work also lays the foundation for a follow-up paper in which we relax the exact \fpo constraint and introduce a metric that quantifies the proximity of an orbital solution to an \fpo.

The paper is structured as follows. In \se{method} we introduce the concept of \fpo and our solution strategy to find \fpos and propagate \fpos throughout phase space. In \se{fpo-props} we make the case that planet systems that experienced sufficient amount of damping, would settle into \fpos and show how our \fpos compare to other works studying resonant interaction. \se{pofit} then fits \fpos to publicly available astrometric and RV data of \pds, \gj, and \hr.

\section{Method} \label{sec:method}
\subsection{Preliminaries and Definitions}
We will number planets from the innermost planet outward with labels $1,2,\ldots,n_p$ where $n_p$ is the number of planets in a resonance chain. Let the chain be described by a sequence of decreasing integers $\bm{N}=N_1{:}N_2{:}...{:}N_{n_p}$ where $N_i{:}N_{i+1}$ is the ratio of the resonance integers involving two such planets. So $\bm{N}$ for the 2:1 chain would simply be 2:1 while $\bm{N}=\textrm{8:4:2:1}$ for the Laplace resonance involving 4 planets. For the 3:2--3:2 resonance, $\bm{N}=\textrm{9:6:4}$. We refer to the first number in $\bm{N}$ as the winding number of the resonant chain.

The separation between two planets $i$ and $i+1$ in resonance is described in terms of the ratio of their instantaneous (osculating) orbital periods, normalized to the period ratio at exact resonance:
\begin{equation}
	\Delta_i \equiv \frac{P_{i+1}/P_i}{N_i/N_{i+1}} -1.
\end{equation}
For a multi-planet system, orbital elements, and by extension $\Delta_i$, osculate. We will denote orbit-averaged quantities by $\langle ... \rangle$.
In particular, $\Delta_1$ evaluated at the start of the \fpo cycle, which is when the innermost body is at periapsis, is referred to as the offset of the resonant chain, $\Delta_1^\dagger$. Along with the planet masses, $\Delta_1^\dagger$ is a grid parameter for the \fpo orbital solutions.

All integrations are carried out with the \rebound simulation package \citep{ReinLiu2012} and we follow \rebound's standards unless mentioned otherwise. In particular, we use Jacobi coordinates (rather than stellar-centric coordinates) with bodies added to the simulation from the innermost planet outwards. Calculations are conducted in the plane. We choose units in which $G=m_\star=1$.

\subsection{Periodic Orbits and coordinates}
\begin{table*}
	\caption{\label{tab:coordinates}Fixed-point orbit coordinates.}
	\centering
	\begin{tabular}{llllp{9.5cm}}
		\hline
		coordinate & symbol                 & dimension  & contains                                     & Meaning                                                                                                              \\ \hline
			grid       & $\bm{x}_\mathrm{grid}$ & $1+n_p$    & $\Delta_1^\dagger, q_1, q_2$ \dots          & resonance offset, planet masses                                                                                      \\[0.8em]
		solution   & $\bm{x}_\mathrm{sol}$  & $4(n_p-1)$ & $e_1, \bm{e}_{21}, \theta_{21},$             & ecc.\ 1$^\textrm{st}$ planet, ecc.~vector 2$^\textrm{nd}$ planet, true longitude difference                          \\
			           &                        &            & $\Delta_2, \bm{e}_{31}, \theta_{31}$ \ldots  & separation, ecc.\ vector, true longitudes of subsequent bodies                                                           \\[0.5em]
			           &                        &            & $t_\mathrm{po}$, $\omega_\mathrm{prec}$      & Complementary solution parameters: the PO period and the system's precession angle after one periodic orbit            \\[0.8em]
			sky        & $\bm{x}_\mathrm{sky}$  & 5          & $t_\mathrm{start}, a_1, \omega_1, i, \Omega$ & time of beginning PO cycle, semi-major axis and argument of periapsis of the innermost body, inclination of the orbital plane, longitude of the ascending node \\
		\hline
	\end{tabular}
	\tablefoot{The coordinates used in this work. Here the eccentricity vector $\bm{e}_{i1}$ is defined with respect to the innermost planet (instead of a reference direction in an inertial frame): $\bm{e}_{i1} = (e_i\cos \omega_{i1}, e_i\sin \omega_{i1})$ where $\omega_{i1} = \omega_i-\omega_1$.}
\end{table*}

Let $\bm{x}$ be a vector containing the quantities describing the orbital state of the system of planetary bodies with the innermost planet at periapsis.  Let $F_N(\bm{x})$ advance this state over $N$ periapsis passages of the innermost body, where $N=N_1$ is the winding number of the resonance chain. An \fpo is an orbit for which this state returns to its starting point:
\begin{equation} \label{eq:po-eq}
	\bm{x} = F_N(\bm{x}).
\end{equation}
In order to find \fpos, one needs to solve this fixed-point equation and define which quantities constitute $\bm{x}$.

\begin{figure}
	\centering
	\includegraphics[width=0.9\columnwidth]{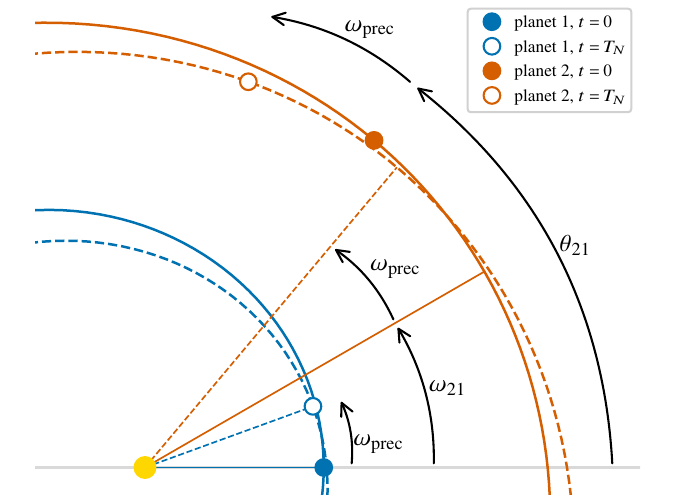}
	\caption{\label{fig:pogeom}Geometry of a fixed-point orbit. Without loss of generality, the innermost planet is at periapsis (indicated by radial lines) at $t=0$. The position and periapsis of a second body are located at an angle $\theta_{21}$ and $\omega_{21}$. After $N$ periapsis passages, corresponding to a time $t_N$, the inner body's periapsis has precessed by an angle $\omega_\mathrm{prec}$. For the system to be periodic, the positions of all other bodies and their periapsis locations should have precessed the same amount, such that the relative angles $\theta_{21}$ and $\omega_{21}$ return to their respective values at $t=0$.}
\end{figure}
To better organize the solution space, we split $\bm{x}$ into discrete grid coordinates (planet masses and resonant chain offset $\Delta_1^\dagger$) and solution coordinates $\bm{x}_\mathrm{sol}$ and solve for the latter, given the former: $\bm{x}_\mathrm{sol} = F_N(\bm{x}_\mathrm{sol}|\bm{x}_\mathrm{grid})$. An overview of the coordinates adopted in this work is given in \Tb{coordinates}.

Which parameters constitute $\bm{x}_\mathrm{sol}$? Without loss of generality we can put the innermost body at periapsis at $t=0$. 
After $N$ passages through periapsis, at time $t_N$, the inner planet again reaches periapsis, although its periapsis direction may have rotated by $\omega_\mathrm{prec}$ (see \fg{pogeom}).
The time $t_N$ is referred to as the period of the \fpo and is also denoted $t_\mathrm{po}$. For the orbital state to repeat, the innermost orbit must also have the same eccentricity as at $t=0$. The same holds for the orbits of the other planets. Each planet's eccentricity vector relative to the innermost body $\bm{e}_{i1} \equiv (e_i\cos \omega_{i1}, e_i\sin \omega_{i1})$ must return to its value at $t=0$. Here and in the following we define $\omega_{i1}=\omega_i-\omega_1$. Further, the orientation of all other planets, again with respect to the innermost body, should be exactly the same at $t=t_\mathrm{po}$ as it is at $t=0$. Here we elect to express this orientation in terms of true longitude, $\theta_{i1} = \theta_i -\theta_1$. Hence, there are $4(n_p-1)$ elements in $\bm{x}_\mathrm{sol}$
	
The choice of $\bm{x}_\mathrm{sol}$ is not unique; for example, one could use Cartesian position and velocity differences relative to the innermost planet \citep{GozdziewskiMigaszewski2020}. We use orbital coordinates for convenience, but express the eccentricity and phase of each outer planet through $\bm{e}_{i1}$ and $\theta_{i1}$. These coordinates remain well behaved as an outer planet's eccentricity approaches zero, when periapsis-based angles become ill-defined. They therefore help solutions propagate across the grid. Still, the construction still relies on a well-defined periapsis of the innermost planet: when $e_1$ is very small, counting its periapsis passages becomes unreliable.

\subsection{The \po data cube} \label{sec:po-cube}
For an $n_p$-planet chain, the \fpo cube has $n_p+1$ dimensions: $n_p$ axes for the planet masses, in addition to $\Delta_1^\dagger$. At each grid point, zero, one or multiple independent solutions may exist. In \app{po-finder} we describe our algorithm to fill the data cube with \fpo solutions, which is essentially based on the premise that neighboring grid points provide good candidate solutions. In this way, several initial solutions can quickly spread through the grid.
We have made these \fpo solutions publicly available (see \app{po-finder}). 

Two caveats should be noted. First, there is no guarantee that \textit{all} solutions have been found; \fpos may have been missed. Second, although we intend to include only stable \fpos in these tables and have conducted a tailored stability analysis for that purpose, their long-term behavior has not been rigorously verified. We are confident that our test (described in \app{po-finder}) ensures stability for the vast majority of cases, but it falls short of a formal stability analysis.

\begin{figure*}
	\includegraphics[width=\textwidth]{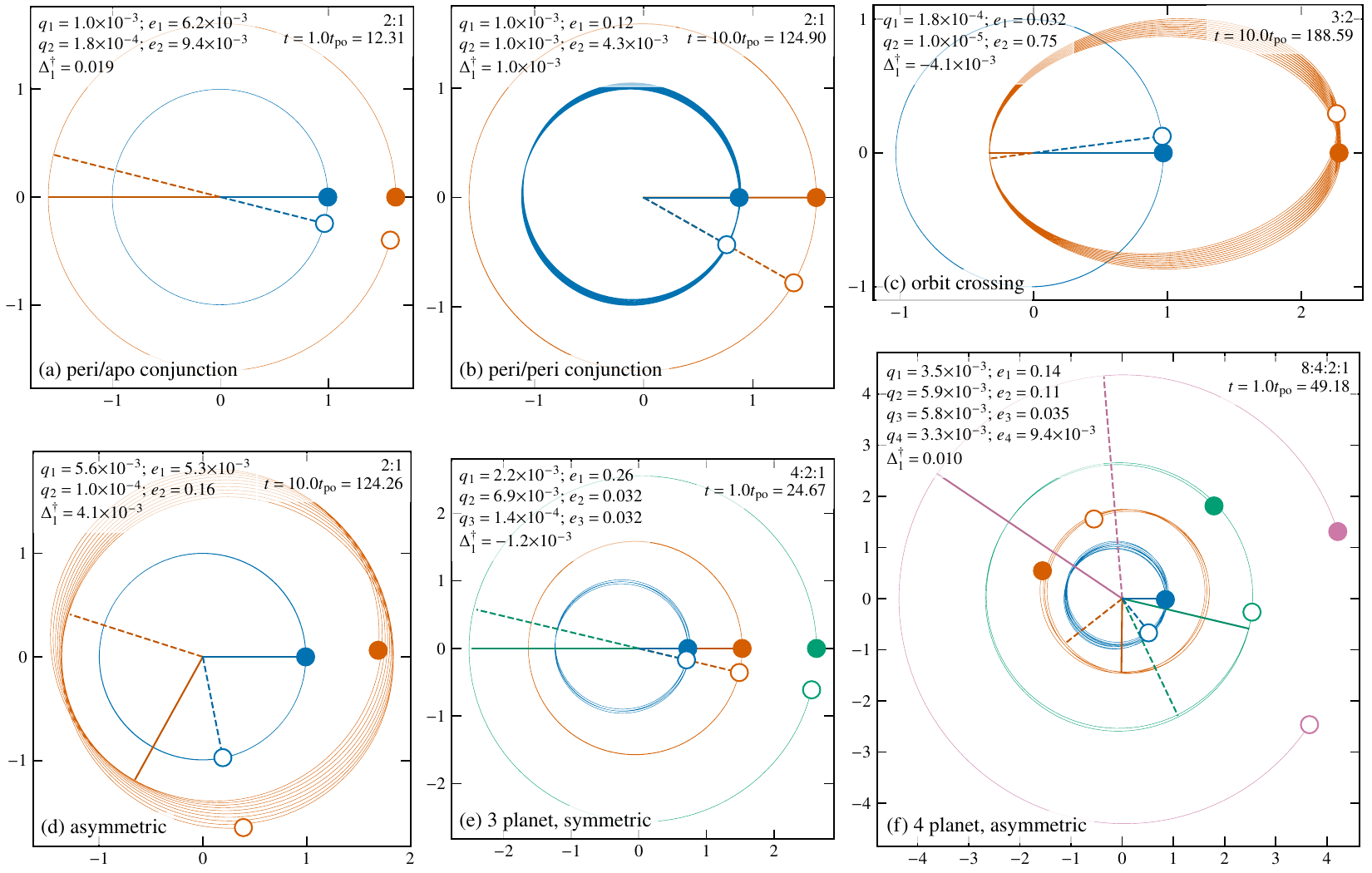}
	\caption{\label{fig:fpos}
        Gallery illustrating the diversity of fPOs. The positions of the planets at the start of an fPO cycle ($t=0$) are indicated by filled circles, and their periapses by solid radial lines. The mass ratios $q_i=m_i/m_\star$ and osculating eccentricities $e_i$ correspond to this initial state. By construction, the innermost planet starts at periapsis (solid blue circle). The system is then integrated for an integer number of fPO periods ($t_\mathrm{po}$), with the final positions and periapses indicated by open circles and dashed lines, respectively. The final configuration matches the initial state modulo a rotation.
    }
\end{figure*}

\section{Properties of \Po} \label{sec:fpo-props}
\subsection{Geometry of fPOs} \label{sec:fpos}
\Fg{fpos} presents a selection of \po solutions found using the grid-search strategy described in \se{method}. By convention, a \po cycle starts with the inner planet at periapsis along the positive x-axis. Panel (a) presents a typical configuration for the 2:1 resonance, in which the planets are at conjunction at $t=0$ with the outer planet at apoapsis. This \fpo is symmetric: the planets lie on a single line at the start of the cycle. After an integration time corresponding to slightly fewer than two inner orbital periods, or slightly less than one outer orbital period, the planets return to exactly the same configuration as at $t=0$, apart from a rotation through $\omega_\mathrm{prec}$. The configuration at any later time can therefore be reconstructed by repeating the cycle and applying the accumulated rotation.

The periapsis--apoapsis configuration at conjunction in panel (a) provides the standard textbook geometrical explanation for resonant stability. However, in the 2:1 resonance, conjunctions can also occur with both planets at periapsis, see panel (b). Because in this example the planets are closer to exact commensurability ($\Delta_1^\dagger \approx 0 $), the precession rate is considerably smaller (the endpoint is now chosen after 10 \fpo cycles). \Po can also be found at higher eccentricity, where the orbits often cross (panel c).  
Because the number of \po solutions grows rapidly at high eccentricity \citep[e.g.,][]{WangMalhotra2017}, we often impose eccentricity limits on the \po cubes. 
 Finally, panel (d) illustrates an asymmetric \po: at $t=0$, the inner planet is at periapsis, but the outer planet is at neither periapsis nor apoapsis.

The discussion naturally extends to multiplanet systems. In panel (e), three planets undergo a triple conjunction in a symmetric \po configuration, while panel (f) shows a four-planet system in an asymmetric \po configuration with a particularly high $\omega_\mathrm{prec}$. In fact, the configurations in the bottom row correspond to our best \po solutions for the \pds, \gj and \hr systems.
\begin{figure*}
	\includegraphics[width=\columnwidth]{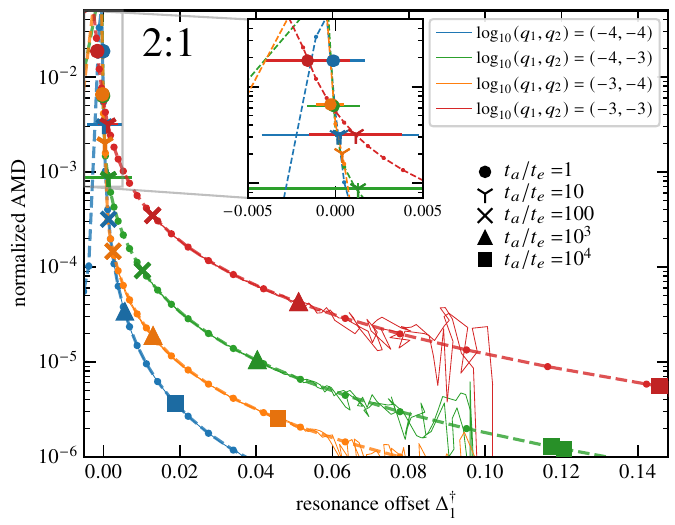}
	\includegraphics[width=\columnwidth]{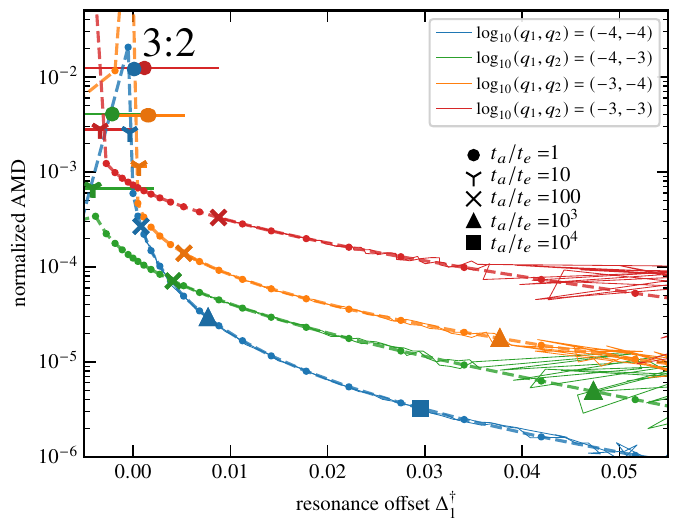}
	\caption{\label{fig:amd}
		Normalized angular momentum deficit (NAMD) for the 2:1 (left) and 3:2 (right) MMRs. Small dots connected by dashed lines show the NAMD of the \po grid solutions for four different mass pairs (colors). The larger symbols show the final states obtained from the N-body simulations, in which the planets are trapped into resonance for different values of $t_a/t_e$. Where significant, the range of libration in $\Delta_1^\dagger$ is indicated by a horizontal line. For the $t_a/t_e=10^2$ simulation, which terminates at the ``$\times$'' symbol, the thin solid line traces the evolution of the NAMD during the simulation. For the N-body simulations, the NAMD is evaluated at the onset of each PO cycle, consistent with the definition of $\Delta_1^\dagger$ used for the \po grid solutions. For sufficiently damped systems, the simulated trajectories follow the \po solution curves in the NAMD–$\Delta_1^\dagger$ plane.}
	\end{figure*}
\begin{figure*}
	\centering
	\includegraphics[width=\textwidth]{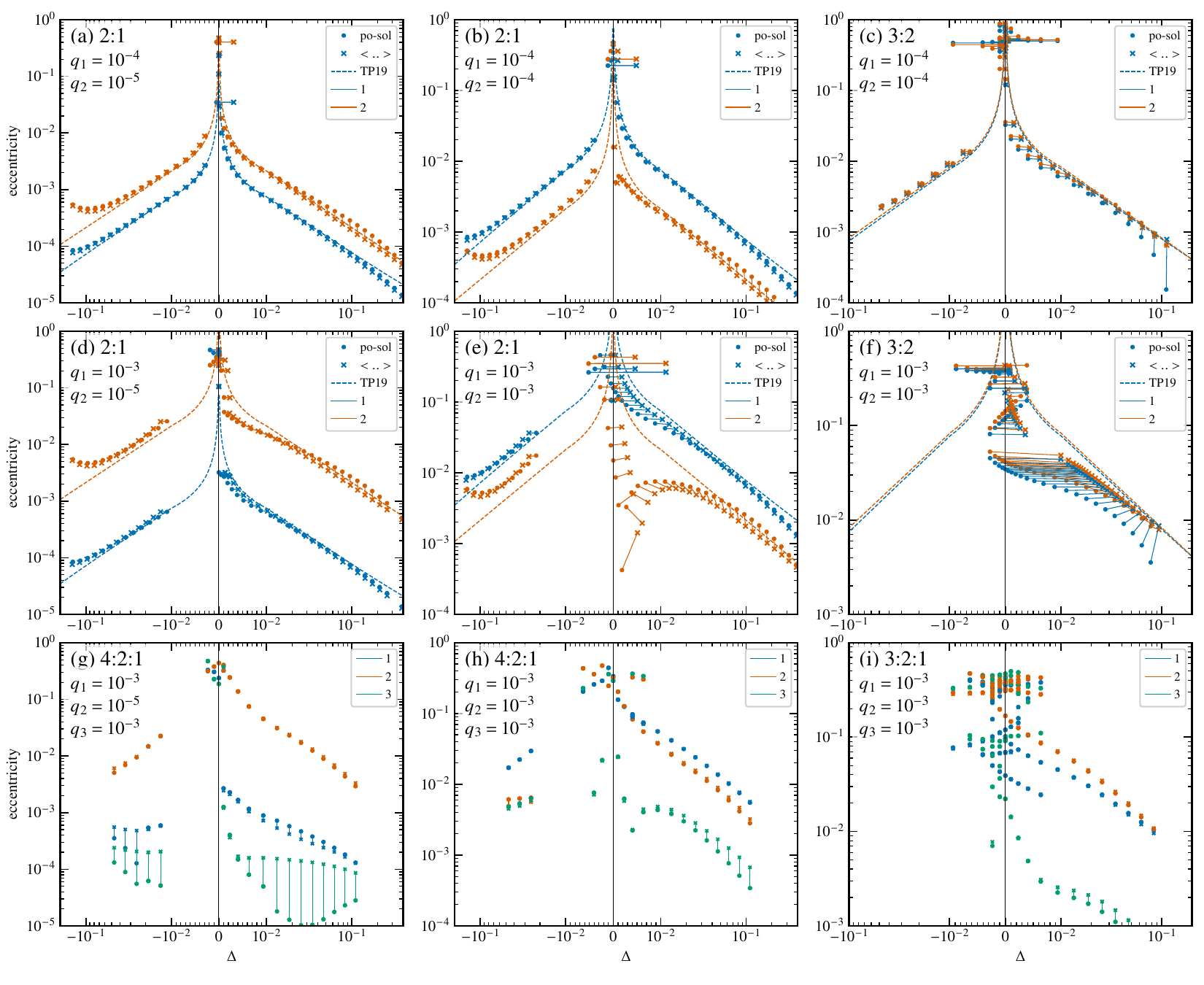}
	\caption{\label{fig:eccplot21}Eccentricity vs resonant offset $\Delta_1$ for the 2:1, 3:2, 4:2:1, and 3:2:1 resonances for several combinations of the two planet masses. The dots show the osculating eccentricities and resonant offsets at the start of each \po cycle, when $\Delta_1=\Delta_1^\dagger$, while the crosses show the corresponding orbit-averaged quantities. The predicted eccentricities according to the \citet{TerquemPapaloizou2019} model are indicated by dashed lines. In the bottom plots, eccentricities are plotted at $\Delta=\Delta_1^\dagger$.}
\end{figure*}

\subsection{The Angular Momentum Deficit} \label{sec:amd}
We loosely interpret \pos as resonant orbits that are maximally damped for a given $\Delta_1^\dagger$. Conversely, we expect \pos to coincide with a state where the Angular Momentum Deficit (AMD) is minimized.

To test this conjecture, we conduct N-body experiments using \rebound with dissipative forces implemented through \texttt{Reboundx} \citep{TamayoEtal2020i}. In each experiment, a planet pair is initialized just wide of either the 2:1 or the 3:2 resonance. The outer planet's semi-major axis is damped on a timescale $t_a$, causing it to converge towards the inner planet and to become trapped in resonance. The eccentricities of both planets are damped on a timescale $t_e$. At $t=0.2t_a$, both the eccentricity and semi-major axis damping are gradually removed over a time $0.4t_a$, such that by $t\simeq0.6 t_a$ the system evolves under purely gravitational dynamics. We adopt $t_a=10^5$ and $t_a=10^6$. In \rebound's default units, the orbital period at the initial location of the inner planet, $a=1$, is $2\pi$. These timescales are mostly sufficiently long to prevent resonance breaking by overstability \citep{LinEtal2025}.

Finally, in the interval $t\in[0.6t_a,\,0.8t_a]$, at the start of each \fpo cycle, the normalized AMD (NAMD) is calculated:
\begin{equation}
	\mathrm{NAMD}
	=
	\frac{
		\displaystyle\sum_{i=1,2} m_i\sqrt{a_i}
		\left(1-\sqrt{1-e_i^2}\right)
	}{
		\displaystyle\sum_{i=1,2} m_i\sqrt{a_i}
	}.
\end{equation}
The results for the 2:1 and 3:2 resonances are presented in \fg{amd} (large symbols). Any libration of $\Delta_1^\dagger$ during this interval is indicated by a horizontal bar (this happens particularly towards high NAMD). For each mass combination (color), symbols could appear twice, because we conduct simulations at two values of $t_a$. However, in many cases the symbols overlap.

In addition, the thin solid lines in \fg{amd} show the \textit{evolution} towards the final state for a damping ratio $t_a=10^2 t_e$. This evolution starts on the right, wide of exact resonance, at low NAMD, since the planets are initialized with 0 eccentricity. 
As the planets converge, $\Delta_1^\dagger$ decreases and the NAMD rises. The cross marks the final trapped state for $t_a/t_e=10^2$.
Any oscillations of NAMD around the corresponding \po curve that were present initially have decayed by this time. The NAMD increases with higher planet mass (color), reflecting stronger resonance forcing.

For comparison, the dashed thick lines show the NAMD of the \po grid solutions obtained in \se{method}.  For simplicity, when multiple \po solutions exist at a given $\Delta_1^\dagger$, we show the solution with the lowest NAMD. We observe that when $t_e<t_a$ the system settles into a trapped state whose NAMD is consistent with that of the \po at the corresponding resonant offset $\Delta_1^\dagger$.

For simulations with weaker eccentricity damping, e.g., $t_a=1t_e$, the situation is somewhat different. In this case, the planets keep converging towards smaller $\Delta$ and the increased resonant forcing drives significant librations in quantities such as $\Delta$, although the NAMD varies much less. The range of $\Delta_1^\dagger$ sampled during the final recording interval is indicated by the horizontal line. Consequently, an arbitrary snapshot of the system is unlikely to correspond exactly to an \po solution. 
Nevertheless, the mean state lies near the \po curve in the NAMD–$\Delta_1^\dagger$ plane, particularly for the 2:1 resonance, suggesting that an \po remains a good approximation. 

A match in NAMD alone does not imply that all components of the solution vector $\bm{x}_\mathrm{sol}$ agree with the \po solution. We checked its components and found that, under sufficiently strong eccentricity damping, the final state approaches the corresponding \po solution.
Finally, we have also performed experiments with only eccentricity damping, in which resonant systems dissipate energy while preserving angular momentum. In these cases, the systems expand: as $\Delta_1$ increases, NAMD decreases, again following the \po solution curve.

We therefore conclude that resonant systems subjected to sustained eccentricity damping will evolve towards \po orbits. Dissipation mechanisms are, of course, also required to bring planet systems into resonance in the first place. Although subsequent (external) perturbations may displace a system from its \po state, the fact that dissipation both establishes resonance and drives the system towards an \po makes \pos a natural blueprint for modelling resonant exoplanet systems.

\subsection{Eccentricity-mass-period offset relationship}
\Fg{eccplot21} illustrates the \po solutions in terms of eccentricity vs period offset for several resonances and planet masses. For each mass combination, we show both solution quantities corresponding to the \po initial state (e.g., $e$ at periapsis) and time-averaged quantities ($e$ averaged along the \po orbit). This distinction is generally small for low masses, but becomes noticeable at higher masses. The same holds for the resonant offset (we plot either $\Delta_1^\dagger$ or $\langle \Delta \rangle$).

At low mass, the $e$--$\Delta$ curves are mostly symmetric about $\Delta=0$. Negative values of $\Delta$ are usually not associated with resonances, because resonance trapping requires that the system converges towards resonance.
At higher masses, some of these solutions tend to become unstable, because the conjunction distance for $\Delta<0$ is closer than for $\Delta>0$. 
Another striking feature is the pileup of high-eccentricity solutions near $\Delta=0$. This is not only due to our finer spacing around $\Delta_1^\dagger=0$; in addition, the multiplicity of the solutions at a given node tends to increase for higher eccentricity as previously noted.  Typically, the low- and high-eccentricity branches are separated by a "gap" in $e$ centered at $e\sim0.1$ (e.g., panels c and d). Since the high-eccentricity solutions are not the focus of our work, we have applied caps on $e$ in many of the \fpo searches, i.e., when $e>e_\mathrm{lim}$, solutions are not included and are not used as candidates for other solutions (see \app{po-finder}).

It is worthwhile to compare our numerical \po solutions with analytical expectations based on the orbit-averaged expansion of the perturbation function, i.e., the Laplace-Lagrange equations. To keep the solutions analytically tractable, typically only terms linear in eccentricity are retained. Specifically, we have followed \citet{TerquemPapaloizou2019}, who also assumed that $\cos^2\phi_i=1$ (dashed lines in \fg{eccplot21}). We generally find excellent matches for low masses.
 At higher masses, however, the analytical predictions deviate from the numerical solutions. In particular, for the 2:1 resonance, there is a transition with increasing $\Delta_1^\dagger$ when the conjunction point switches from P/P to P/A (e.g., panel e). In these cases, the resonance angle $\phi_2$ librates and the assumption $\cos^2\phi_2=1$ breaks down. Once eccentricities exceed $0.01$, the predicted and numerical eccentricities can differ by factors of order unity.

Finally, the bottom rows of \fg{eccplot21} show eccentricities for a three-planet system. Here we use the same masses as in the rows above and add a Jupiter-mass third planet ($q_3=10^{-3}$) to the chain. This additional forcing raises the eccentricities of the second planet. 
If an additional outer planet is present but undetected, the outermost observed planet may have a higher eccentricity than predicted by a two-planet \po model.

\begin{table*}
	\centering
	\caption{Range of fitting parameters (sky parameters) for \pds, \gj, and \hr} \label{tab:skypars}
	\begin{tabular}{p{4cm} l r r r l p{3cm}}
		\hline
		quantity                     &                                       & \multicolumn{3}{c}{range} & unit       & comment                             \\
		                             &                                       & PDS 70                    & GJ 876     & HR 8799   &                         \\
		\hline
		resonance chain              &                                       & 2:1                       & 4:2:1      & 8:4:2:1                             \\
		ascending node               & $\Omega$                              & 265--300                  & 0          & 0--180    & deg                     \\
		argument of periapsis        & $\omega_\mathrm{inner}$               & 0--$2\pi$                 & 0--$2\pi$  & 0--$2\pi$ & rad                     \\
		sky gravitational constant   & $\tilde{G}$                           & 0.17--0.22                &            & 5.8--7.2  & mas$^3$day$^{-2}$ & (a) \\
		stellar mass                 & $m_\star$                             &                           & 0.37       &           & $m_\odot$         & (c) \\
		semi-major axis of the inner planet & $a_\mathrm{inner}$                    & 180--220                  & 0.13--0.14 & 370--410  & mas (au)          & (c) \\
		inclination                  & $i$                                   & 48--55                    & 120--130   & 148--162  & deg                     \\
		\po epoch                    & $\varphi_\mathrm{po}(t_\mathrm{ref})$ & 0--1                      & 0--1       & 0--1      &                   & (b) \\
		\hline
	\end{tabular}
	\tablefoot{The parameter ranges are motivated by \citet{KepplerEtal2019,TrevascusEtal2025} for \pds, \citet{MillhollandEtal2018} for \gj, and \citet{ZurloEtal2022} for \hr.
		(a) See \eq{Gtilde} for the definition of $\tilde{G}$.
	        (b) The \po epoch is specified by the \fpo phase $\varphi_\mathrm{po}(t_\mathrm{ref})$ at the reference time $t_\mathrm{ref}$, the time of the first RV or astrometric observation. Equivalently, it can be specified by $t_\mathrm{start}$, the time at which the \po cycle begins ($\varphi_\mathrm{po}=0$).
		(c) For GJ 876, only radial velocity data are modeled, which are insensitive to the distance to the system. The stellar mass is used instead of the parameter $\tilde{G}$, and $a_\mathrm{inner}$ is expressed in units of au.}
\end{table*}

\section{Fitting \fpos to astrometric and RV data} \label{sec:pofit}

We fit \po to archival astrometric and radial velocity data of the \pds, \gj, and \hr multiplanet systems (\citealt{MillhollandEtal2018,ZurloEtal2022,TrevascusEtal2025}; summarized in \app{astromdata}). These systems all show clear indications of commensurabilities in the planets' period ratios, although only for \gj has a resonance structure clearly been established \citep{RiveraEtal2005}. 
These fits demonstrate that the \po method can reproduce the data with plausible orbital solutions and provide starting points for more detailed characterization of each system.

Expressed in angular units ($r\rightarrow r/d$), Newton's equation of motion reads:
\begin{equation}
	\label{eq:eom}
	\ddot{\bm{r}}_i
	= -\frac{Gm_\star}{d^3} \sum_{j\neq i} q_j\frac{\bm{r}_j-\bm{r}_i}{|\bm{r}_j-\bm{r}_i|^3}
	\equiv -\tilde{G} \sum_{j\neq i} q_j\frac{\bm{r}_j-\bm{r}_i}{|\bm{r}_j-\bm{r}_i|^3}
\end{equation}
where $d$ is the distance to the planetary system, and $q_i=m_i/m_\star$ the planet-to-stellar mass ratio. Expressing $r_i$ in units of milliarcseconds (mas) and time in units of days (d), the gravitational constant $\tilde{G}$ in \eq{eom} becomes:
\begin{align}
	\label{eq:Gtilde}
	\tilde{G}
	 & = \frac{Gm_\star}{d^3} \times \left( \frac{\mathrm{mas}}{\mathrm{rad}} \right)^3 \left( \frac{\mathrm{d}}{\mathrm{sec}} \right)^2
	= \frac{Gm_\star}{\mathrm{au}^3} \left( \frac{p}{\mathrm{mas}} \right)^3                                                             \\ 
	 & = 2.959\times10^{-4} \left( \frac{m_\star}{m_\odot} \right) \left( \frac{p}{\mathrm{mas}} \right)^3 \mathrm{mas^3\ d^{-2}}
\end{align}
where $p$ is the parallax. \Eq{Gtilde} shows that when fitting astrometric data, the stellar mass and distance (or parallax) combine into a single quantity $\tilde{G}$, which acts as a modified gravitational constant. We consider $\tilde{G}$ as a parameter instead of separately fitting for $d$ and $m_\star$. In \rebound, the value of the gravitational constant is set accordingly.

\begin{table*}[t]
	\centering
	\caption{Selected grid search results for \pds.}
	\label{tab:pds70cid}
	\begin{tabular}{lrrr|rr|rrrrr|rr}
id & $\Delta_1^\dagger$ & ${m}_\mathrm{b}$ [$m_\star$] & ${m}_\mathrm{c}$ [$m_\star$] & ${e}_\mathrm{b}$ & ${e}_\mathrm{c}$ & $a_\mathrm{b}$ [mas] & $\omega_\mathrm{b}$ [$^\circ$] & $i$ [$^\circ$] & $\Omega$ [$^\circ$] & $\tilde{G}$ & $\chi^2$ & $\chi^2_\mathrm{optm}$ \\
\hline
$ 1$ & $4.1{\times}10^{-3}$ & $0.00562$ & $1.00{\times}10^{-4}$ & $0.005$ & $0.156$ & $199.02$ & $165.3$ & $54.2$ & $282.1$ & $0.2253$ & $107$ & $107$ \\
$ 2$ & $1.0{\times}10^{-3}$ & $0.00178$ & $1.00{\times}10^{-4}$ & $0.016$ & $0.163$ & $197.19$ & $166.2$ & $54.2$ & $282.4$ & $0.2236$ & $108$ & $107$ \\
$ 3$ & $6.3{\times}10^{-3}$ & $0.0100$ & $3.16{\times}10^{-4}$ & $0.010$ & $0.167$ & $198.48$ & $166.4$ & $54.5$ & $282.4$ & $0.2276$ & $109$ & $107$ \\
$ 4$ & $2.3{\times}10^{-3}$ & $0.00562$ & $5.62{\times}10^{-4}$ & $0.030$ & $0.169$ & $194.18$ & $165.2$ & $53.8$ & $283.0$ & $0.2257$ & $109$ & $107$ \\
$ 5$ & $2.3{\times}10^{-3}$ & $0.0100$ & $0.00178$ & $0.054$ & $0.161$ & $190.30$ & $160.1$ & $52.6$ & $283.4$ & $0.2193$ & $110$ & $107$ \\
$ 6$ & $9.4{\times}10^{-3}$ & $1.78{\times}10^{-4}$ & $0.0100$ & $0.183$ & $0.001$ & $196.09$ & $106.6$ & $50.1$ & $290.7$ & $0.1852$ & $110$ & $110$ \\
$ 7$ & $9.4{\times}10^{-3}$ & $0.00100$ & $0.0100$ & $0.178$ & $0.002$ & $195.93$ & $107.1$ & $50.2$ & $290.6$ & $0.1863$ & $110$ & $110$ \\
$ 8$ & $-1.0{\times}10^{-3}$ & $0.0100$ & $0.00316$ & $0.095$ & $0.149$ & $185.43$ & $152.2$ & $51.2$ & $284.5$ & $0.2094$ & $110$ & $109$ \\
$ 9$ & $2.3{\times}10^{-3}$ & $0.00100$ & $0.00316$ & $0.167$ & $0.007$ & $194.97$ & $101.4$ & $49.9$ & $290.7$ & $0.1868$ & $111$ & $110$ \\
$10$ & $0.019$ & $1.00{\times}10^{-4}$ & $0.0178$ & $0.185$ & $0.000$ & $196.55$ & $113.1$ & $50.6$ & $290.3$ & $0.1862$ & $111$ & $111$ \\
\hline
\end{tabular}

	\tablefoot{A selection of PO solutions fitted to the PDS 70 system. The first entry denotes the best-fit solution of all grid points. Columns denote grid parameters (principal separation and planet masses), the eccentricities of the corresponding \po solution, several fitted sky parameters, and the $\chi^2$ value at the grid point. The final column gives $\chi^2$ after the solution is allowed to move away from the grid point. See \Tb{skypars} for a description of the sky parameters. Masses are in units of the stellar mass. The ${\rightarrow}X$ in the $\chi^2_\mathrm{optm}$ column indicates that the solution converges to the same minimum as in the $\textrm{id}=X$ row.}
\end{table*}

Therefore, if \textit{only} astrometric data is fitted (here: \pds and \hr), $p$ and $m_\star$ are degenerate through $\tilde{G}$. Velocities would be expressed in units of mas\,day$^{-1}$ in our system, but we do not use them. Conversely, if only RV data is available, we can place the system at a nominal distance of 1\,kpc (corresponding to a 1\,mas parallax) so that distances are expressed in au and velocities in physical units. If both RV and astrometric data are present (not in this work), no such degeneracies exist: the solution requires a distance scale ($p$) as well as a stellar mass.

In an \fpo, planets orbit in the same plane. The \fpo fixes the innermost planet's eccentricity $e_1$ and, for each outer planet, its eccentricity $e_i$, relative argument of periapsis $\omega_{i1}$, resonance offset $\Delta_i$, and relative true longitude $\theta_{i1}$ at the start of the \fpo cycle.

There are six remaining parameters that determine the position of the system's planetary motion on the sky:
\begin{enumerate}
	\item The argument of the ascending node $\Omega$. This angle is measured in the direction from RA to declination, i.e., clockwise on the sky (See \fg{coordsystem}).
       \item The inclination $i$ of the orbital plane. The orbit is face-on at $i=0^\circ$ and $180^\circ$ and edge-on at $i=90^\circ$. In our convention, the apparent motion as seen on the sky is clockwise for $i<90^\circ$ and anti-clockwise for $i>90^\circ$.
	\item The start of the present PO cycle $t_\mathrm{start}$. This should lie before the time of the first data point (which is usually referred to the reference time $t_\mathrm{ref}$). The PO solution is initialized at $t=t_\mathrm{start}$.
	\item The semi-major axis (at $t_\mathrm{start}$) of the innermost planet $a_1$;
	\item The argument of periapsis $\omega_1$ of the innermost planet at $t_\mathrm{start}$.
	\item The value of the modified gravitational constant $\tilde{G}$.
\end{enumerate}
In the case of RV-only data, $\tilde{G}$ can be replaced by a fixed stellar mass $m_\star$.

Given an \fpo solution, these six parameters completely specify the motion on the sky. The number of these sky parameters is independent of the dimensionality of the problem (i.e., the number of planets in the system).

We characterize the fidelity of the solution by the goodness-of-fit parameter $\chi^2$:
\begin{equation}
	\chi^2 = \sum_i \left( \frac{\Delta \mathrm{RA}_i}{\sigma_{\mathrm{RA},i}} \right)^2
	+ \left( \frac{\Delta \mathrm{dec}_i}{\sigma_{\mathrm{dec},i}} \right)^2
	+ \left( \frac{\Delta V_{\mathrm{z},i}}{\sigma_{\mathrm{rv,z},i}} \right)^2 .
\end{equation}
Here $\Delta \mathrm{RA}_i$ ($\Delta \mathrm{dec}_i$ and $\Delta V_{\mathrm{z},i}$) are the model residuals, and each $\sigma$ is the uncertainty of the corresponding measurement.

Our $\chi^2$ minimization consists of two stages: on-grid and off-grid.
First, for a given solution at a grid point in the \po cube (\se{po-cube}), we aim to find the sky parameters ($\bm{x}_\mathrm{sky}$) that minimize $\chi^2$. This is a global minimization problem and because $\chi^2(\bm{x})$ is possibly strongly multimodal, optimization is not trivial, even with the vastly reduced number of parameters from the \fpo solutions. We adopt the covariance matrix adaptation evolution strategy (CMA-ES; \texttt{cma 4.4.3}; \citealt{HansenEtal2019}) to find the global minimum at every grid point. Because some of these solutions may miss the minimum, we "refine" the solutions in the following way. First, we select those grid points that have low $\chi^2$ values (e.g., within a certain factor of the global minimum). We then rerun the CMA-ES search on those grid points \textit{and their neighboring} points several more times. These repeated searches reduce the chance that the optimization misses the lowest $\chi^2$ value at a grid point.

Second, we relax the condition that solutions are restricted to grid points. For a selected number of low-$\chi^2$, on-grid solutions, an additional optimization is performed using the Levenberg-Marquardt (LM) algorithm \citep{PressEtal2007}. Specifically, we minimize $\chi^2 (\bm{x})$ where $\bm{x} = \{\bm{x}_\mathrm{grid}, \bm{x}_\mathrm{sky}\}$, treating $\bm{x}_\mathrm{grid}$ as continuous rather than discrete and updating the solution parameters $\bm{x}_\mathrm{sol}$ on the fly, according to the current value of $\bm{x}_\mathrm{grid}$. Only steps that decrease $\chi^2$ are accepted. We terminate the optimization when no further decrease is achieved or when the solution leaves the computational domain (e.g., when one of the masses becomes too small).

\begin{figure*}
	\sidecaption
	\includegraphics[width=0.7\textwidth]{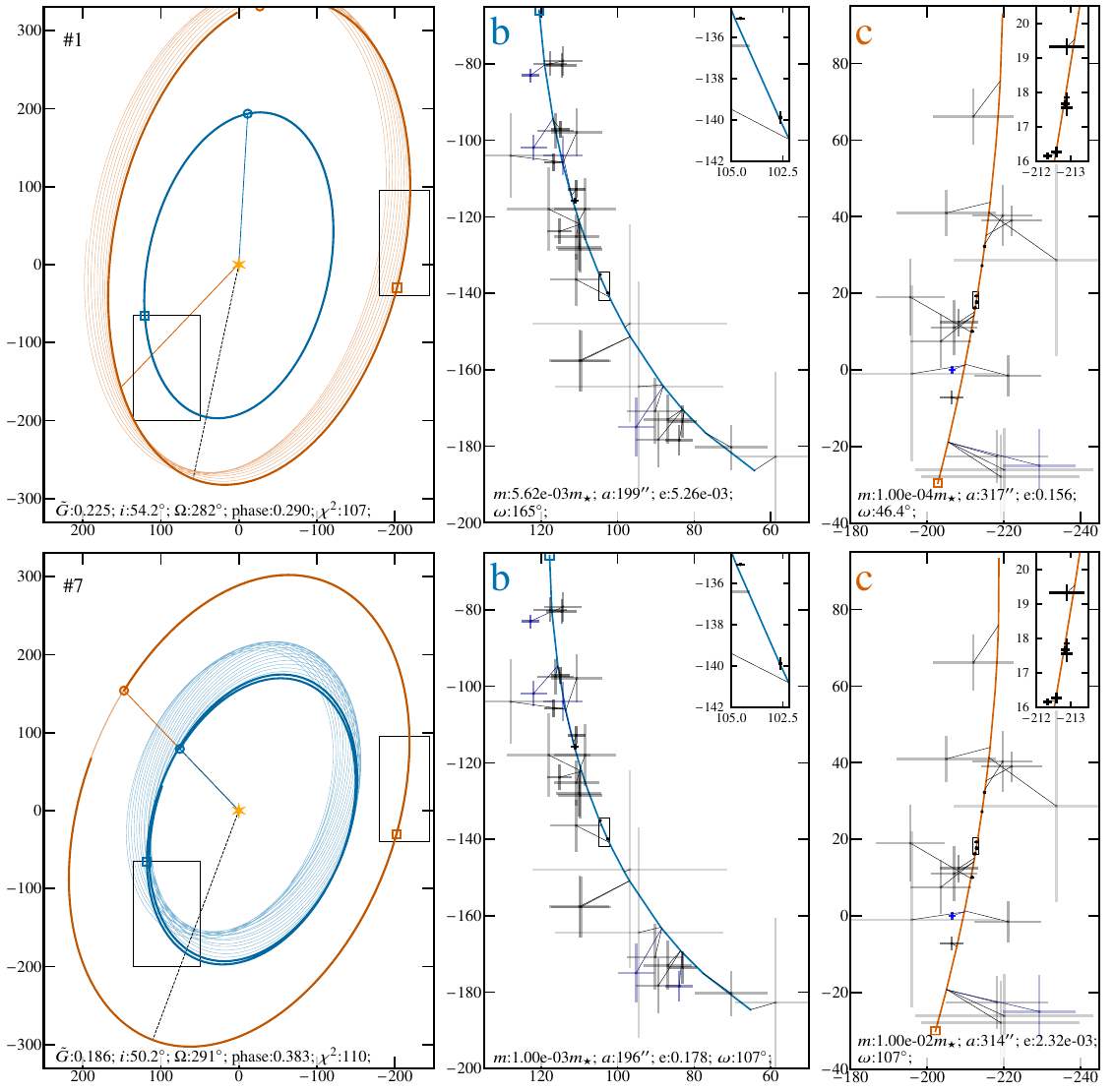}
	\caption{\label{fig:pds70}\label{fig:pds70-alt}Two \fpo fits to the astrometric data for the PDS 70 system (\#1 and \#7 of \Tb{pds70cid}). The left overview panels show one full \fpo in bold. At the start of the PO cycle ($t=t_\mathrm{start}$), circles indicate the initial positions of the planets, while radial lines mark the periapses of their orbits. Thin lines show the subsequent orbital evolution, illustrating the precession of their orbits. The radial black dashed line indicates the ascending node ($\Omega$), and squares indicate the present-day positions of the planets. Panels (b) and (c) show the orbital solution together with the astrometric data for the respective planets.
								The upper orbit is an example of Group I (more massive planet b and eccentric planet c) while the lower fit belongs to Group II (more massive c and eccentric b).  Despite the different orbital parameters, the astrometric fits are nearly identical, although the present-day position of planet b (blue square) lies slightly farther inward in the Group II example.}
\end{figure*}
\subsection{PDS 70} \label{sec:pds70}
The sampled ranges in the sky-plane parameters are listed in \Tb{skypars}. For PDS 70, the adopted ranges for $\Omega$ and $i$ are based on the values reported for the parent circumstellar disk \citep{KepplerEtal2019}. Likewise, the ranges for $a_\mathrm{inner}$ and $\tilde{G}$ are loosely based on the study of \citet{TrevascusEtal2025}, although we adopt considerably broader bounds. No constraints are imposed on the remaining two sky-plane parameters, the argument of periapsis of planet b, $\omega_b$, and the epoch of its periapsis passage $t_\mathrm{start}$.

The best-fitting \fpo models are listed in \Tb{pds70cid}. The solutions are ordered by increasing $\chi^2$, subject to the requirement that adjacent solutions are separated by at least two grid points and that all planet masses exceed $m/m_\star=10^{-4}$. Broadly, the solutions fall into two groups. Group I (\#1-5 in \Tb{pds70cid}), of which the top panel in \fg{pds70} shows an example, has a higher mass for planet b than for planet c, a high $\tilde{G}$ and a planet c eccentricity of ${\approx}0.16$. Its underlying \fpo is asymmetric, with resonant angles librating at values other than 0 or $\pi$. Group II (\#6, 7, 9, and 10), illustrated by the bottom panel in \fg{pds70-alt}, has a lower $\tilde{G}$ (read: stellar mass), a lower planet b mass, and a higher planet c mass. Its underlying \fpo is symmetric, with both resonance angles librating around 0. Solution \#8 is an exception and does not belong to any of these groups. Although the difference in $\chi^2$ between the two groups may suggest a preference for Group I over Group II, potential correlations among clustered data points warrant some caution in interpreting this difference.  Allowing the solutions to move away from the grid points improves the fit only marginally ($\chi^2_\mathrm{optm} \approx \chi^2$; last column of \Tb{pds70cid}). The presence of several distinct solutions, combined with the small improvements obtained by moving off the grid, suggests a parameter space landscape characterized by multiple local minima. Additional high-precision astrometric data are therefore required to further constrain the orbital architecture.

Adopting the parallax reported by \citet{GaiaCollaborationEtal2021}, $p=8.8975\,\mathrm{mas}$, and neglecting its 0.21\% uncertainty allows us to convert $\tilde{G}$ and $a_b$ into physical units. A $1\,m_\odot$ star then corresponds to $\tilde{G}=0.2084$, and our solutions imply a stellar mass ranging from $0.89\,m_\odot$ (Group II) to $1.08\, m_\odot$ (Group I). The lower-mass solutions (typically Group II) are therefore more consistent with the mass inferred from stellar spectral fitting ($m_\star = 0.76{-}0.88\, m_\odot$) \citep{MuellerEtal2018ii,WangEtal2021ii}. In addition, the node angle of Group II aligns more closely with the inferred position angle of 157--160$^\circ$ ($\Omega=290{-}293^\circ$) for the \pds disk \citep{KepplerEtal2019}.

Comparing our solutions with the ``Stable (incl.~N-body)'' column of Table 3 in \citet{TrevascusEtal2025}, we generally find good agreement for most parameters, particularly when restricting our results to Group II. Our parameter search nevertheless allows a broad range of values. For example, within Group II, the likelihood-weighted 68\% (95\%) confidence intervals for the masses of planets b and c are $m_b = 0.3^{+0.6(+16)}_{-0.2(-0.28)}\,m_\mathrm{J}$ and $m_c = 9.2^{+7(+20)}_{-4(-8)}\,m_\mathrm{J}$, respectively (for these estimates, we adopted a stellar mass of $0.88\,m_\odot$ and included only solutions with $\tilde{G}<0.20$). These wide ranges reflect the logarithmic sampling of the mass grid, the absence of explicit priors, and, at the upper end, the compatibility of the \fpo models with high planet masses. Nevertheless, the results provide some evidence that planet c is more massive than planet b in the Group II branch.

Future astrometric measurements may distinguish between these two groups. In Group II, planet b is approaching periapsis, so its orbit exhibits stronger curvature and greater inward radial motion than the solutions in Group I. Our simulations already predict an offset of several mas, within the precision of GRAVITY (see the blue square at the top of the planet b panels in \fg{pds70}).

\subsection{\gj (4:2:1)} \label{sec:gj876}
\begin{table*}
	\centering
	\caption{Selected grid search results for \gj.}
	\label{tab:gj876cid}
	\begin{tabular}{lrrrr||rrrr|rr}
id & $\Delta_1^\dagger$ & ${m}_\mathrm{c}$ [$m_\star$] & ${m}_\mathrm{b}$ [$m_\star$] & ${m}_\mathrm{e}$ [$m_\star$] & $a_\mathrm{c}$ [au] & $\omega_\mathrm{c}$ [$^\circ$] & $i$ [$^\circ$] & $\phi$ & $\chi^2$ & $\chi^2_\mathrm{optm}$ \\
\hline
$ 1$ & $0.0$ & $0.00178$ & $0.00562$ & $1.00{\times}10^{-4}$ & $0.137$ & $227.7$ & $120.0$ & $0.5$ & $3.19{\times}10^{4}$ & $2470$ \\
$ 2$ & $1.0{\times}10^{-3}$ & $0.00178$ & $0.00562$ & $5.62{\times}10^{-4}$ & $0.137$ & $289.2$ & $120.0$ & $0.4$ & $6.52{\times}10^{4}$ & $\rightarrow$1 \\
$ 3$ & $-1.0{\times}10^{-3}$ & $0.00178$ & $0.00562$ & $0.00178$ & $0.137$ & $216.0$ & $120.0$ & $0.5$ & $1.08{\times}10^{5}$ & $\rightarrow$1 \\
$ 4$ & $-2.3{\times}10^{-3}$ & $0.00316$ & $0.00562$ & $1.00{\times}10^{-5}$ & $0.137$ & $220.2$ & $126.6$ & $0.5$ & $1.14{\times}10^{5}$ & $9.46{\times}10^{4}$ \\
$ 5$ & $1.0{\times}10^{-3}$ & $0.00100$ & $0.00562$ & $3.16{\times}10^{-4}$ & $0.137$ & $218.3$ & $120.0$ & $1.0$ & $1.26{\times}10^{5}$ & $5340$ \\
$ 6$ & $2.3{\times}10^{-3}$ & $0.00100$ & $0.00562$ & $0.00100$ & $0.137$ & $295.1$ & $120.0$ & $0.4$ & $1.51{\times}10^{5}$ & $5.37{\times}10^{5}$ \\
$ 7$ & $-1.0{\times}10^{-3}$ & $0.00316$ & $0.00562$ & $1.78{\times}10^{-4}$ & $0.137$ & $295.6$ & $127.7$ & $0.4$ & $1.54{\times}10^{5}$ & $\rightarrow$1 \\
$ 8$ & $-4.1{\times}10^{-3}$ & $0.00316$ & $0.00562$ & $0.00178$ & $0.137$ & $216.8$ & $130.0$ & $0.5$ & $1.74{\times}10^{5}$ & $\rightarrow$1 \\
$ 9$ & $0.0$ & $0.00100$ & $0.00562$ & $0.00100$ & $0.137$ & $132.2$ & $120.0$ & $0.1$ & $1.95{\times}10^{5}$ & $1.21{\times}10^{4}$ \\
$10$ & $-2.3{\times}10^{-3}$ & $0.00178$ & $0.00562$ & $0.00178$ & $0.137$ & $119.8$ & $120.0$ & $0.1$ & $2.10{\times}10^{5}$ & $2.70{\times}10^{4}$ \\
\hline
$ 1$ & $-1.2{\times}10^{-3}$ & $0.00217$ & $0.00694$ & $1.38{\times}10^{-4}$ & $0.137$ & $234.5$ & $127.8$ & $0.5$ & $2470$ \\
\hline
\end{tabular}

	\tablefoot{Top: grid search results for the 4:2:1 resonance in \gj, similar to \Tb{pds70cid}. A stellar mass of $0.37\,m_\odot$ has been assumed. The off-grid optimization (last column) significantly improves the fit. The parameters corresponding to the $\chi^2$ minimum thus obtained are listed in the bottom row. Adopting any other value for $m_\star$ will give the same results with $a_c\propto m_\star^{1/3}$ and $\sin i \propto m_\star^{-1/3}$.}
\end{table*}

\begin{figure*}
	\includegraphics[width=\textwidth]{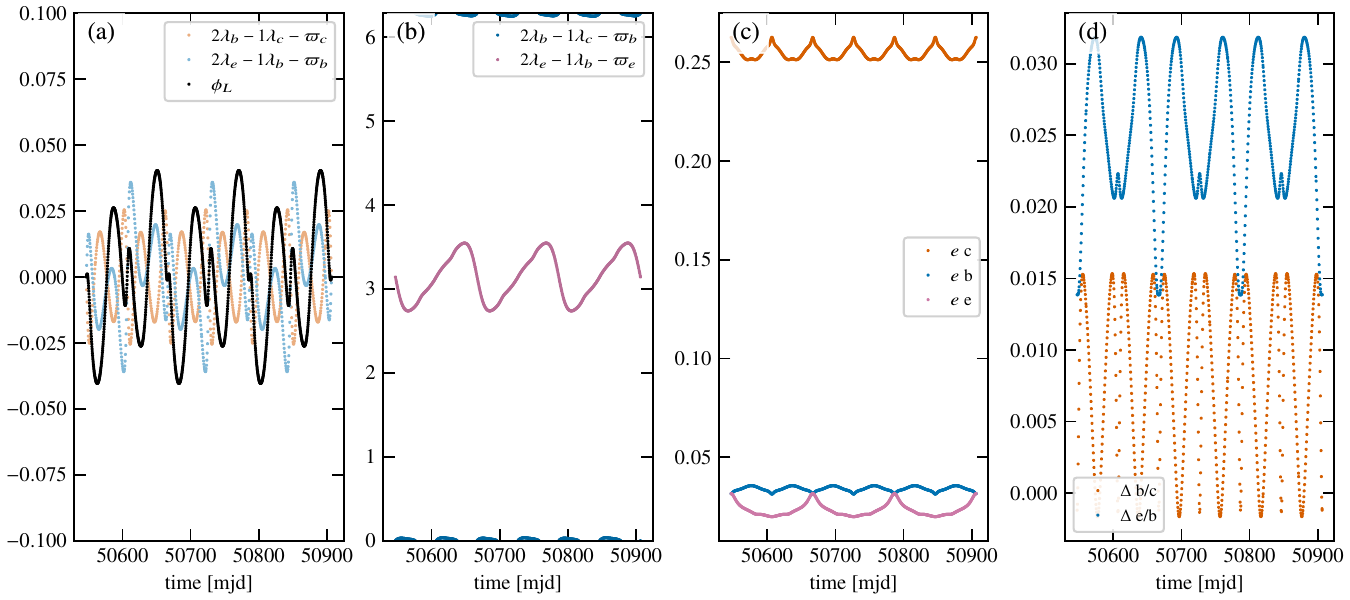}
	\caption{\label{fig:gj876angles}Resonance angles, eccentricities, and period offsets over the course of three \po periods ($3t_\mathrm{po}$) for the best-fit \fpo solution to \gj. (a) Inner two-body resonance angles and the Laplace angle: $\phi_L = 2\lambda_e -3\lambda_b +\lambda_c$. (b) Outer two-body resonance angles. Note the zoom around 0 in panel (a). (c) Osculating eccentricities. (d) Resonance offsets. }
\end{figure*}

\begin{figure*}
	\includegraphics[width=\textwidth]{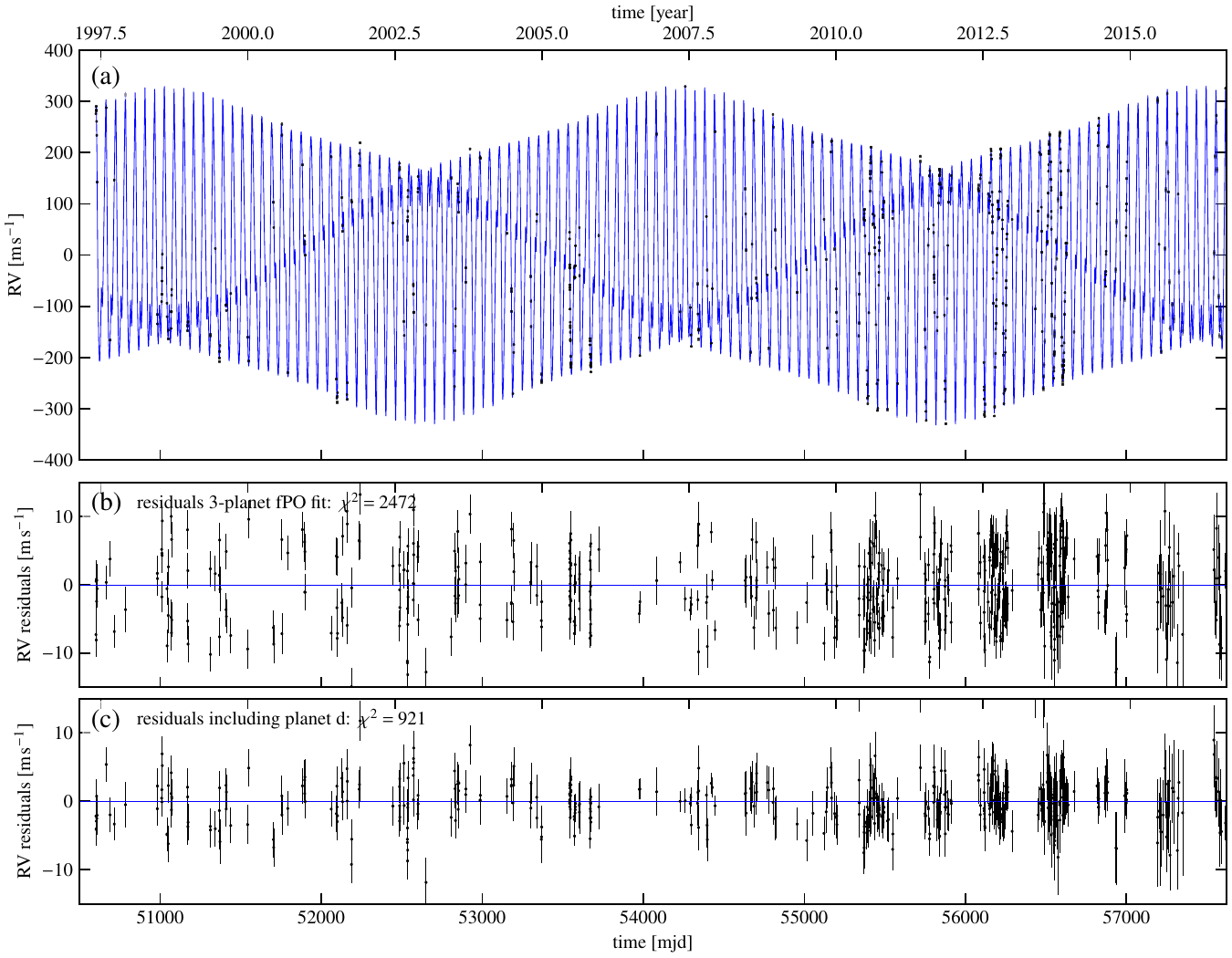}
	\caption{\label{fig:gj876rv}Best fit (a) and residuals (b) for the radial velocity data of \gj. The model follows a strict \po orbit with planets c/b/e in a 4:2:1 resonance. (c) Residuals corrected after subtracting the signal caused by the non-resonant planet d.}
\end{figure*}

\gj is a system where four planets have been detected through radial velocity (RV) measurements. Compared to \pds, \gj is a far more compact system, and a 4:2:1 Laplace resonance among planets c, b, and e has been firmly established \citep{RiveraEtal2010}. We focus on these three resonant planets and, for now, neglect the smaller, innermost planet d. We adopt the RV data compiled by \citet{MillhollandEtal2018} (see \app{gj876data}); that study and others \citep[e.g.,][]{TrifonovEtal2018} have already analyzed the system in depth. Here, we simply assess whether the RV data are consistent with a \po solution and whether the inferred orbital and physical properties agree with those reported in these studies.

The RV observations span roughly 200 orbits of the innermost resonant planet and occur at irregular times. Direct N-body fitting is therefore computationally costly, whereas a Keplerian fit would neglect planet--planet interactions.  
However, under the \fpo assumption, these concerns largely vanish. Because the solution repeats itself, only one full \po spanning ${\approx}$120 days needs to be integrated. From this, all other data points follow:
\begin{equation}
	\bm{X}(t+kt_\mathrm{po}) = [\mathrm{R}(\omega_\mathrm{prec},\bm{n})]^k \bm{X}(t);\qquad k=0,1,\ldots
\end{equation}
Here $\bm{X}$ denotes a position or velocity relative to the system barycenter, $\bm{n}$ is normal to the orbital plane, and $\omega_\mathrm{prec}$ is the rotation angle per \po cycle. The matrix $\mathrm{R}(\phi,\bm{n})$ rotates a vector through angle $\phi$ about $\bm{n}$.
\footnote{$\mathrm{R}(\phi,\bm{n}) = I + \sin\phi\,K + (1-\cos\phi)\,K^2$, where $K_{ij} = -\epsilon_{ijk} n_k$, that is, the unique matrix satisfying $K\bm{x} = \bm{n} \times \bm{x}$ for any vector $\bm{x}$.} 
For an observation at epoch $T$, write $T-t_\mathrm{start}=(k+\phi_\mathrm{po})t_\mathrm{po}$, where $k$ is a nonnegative integer cycle count and $0\leq\phi_\mathrm{po}<1$ is the fractional phase within a cycle. 
The state at $T$ follows by integrating the model to time $t=\phi_\mathrm{po}t_\mathrm{po}$ and applying the rotation in the preceding equation $k$ times.
We found that the strict periodic nature of \fpos reduces the numerical expense by at least a factor of 10.

Because the RV model is independent of parallax, we set the parallax to $1\,\mathrm{mas}$ to express lengths in au. For ease of comparison with previous studies, we also fix the stellar mass at $m_\star=0.37\,m_\odot$, such that our modified gravitational constant becomes $\tilde{G} = Gm_\star = 1.0949\times10^{-4}\, \mathrm{au^3\,day^{-2}}$. RV data are insensitive to the node angle $\Omega$, which we therefore set to zero. Furthermore, in a multi-planet system, the planet-to-star mass ratios will be constrained through the bodies' mutual gravitational interaction. Once these are constrained, the overall amplitude of the RV signal depends on the inclination of the system, which we leave as a free parameter. We adopt the $i>90^\circ$ branch based on HST astrometry \citep{BeanSeifahrt2009}.\footnote{They report an inclination ${<}90^\circ$, which we interpret as anti-clockwise motion on the sky following the standard HST astrometry. Following our convention for the coordinate system (see \fg{coordsystem}), $i>90^\circ$.}

The results of our on- and off-grid $\chi^2$ minimization are presented in \Tb{gj876cid}. The grid minimization produces a diversity of outcomes, characterized by a large range in $\chi^2$, with the algorithm especially constraining the mass of the middle planet b well. In contrast to \pds, the off-grid minimization results in a clear minimum, significantly improving the fit, at $\chi^2 = 2470$---more than tenfold lower than the on-grid minimization. Only about half of the grid nodes converge to this minimum during off-grid optimization, showing the value of multiple starting points. Focusing on our best-fit solution (\fg{gj876angles}), it is clear that the system is deep in resonance ($\Delta\approx0$) with the amplitude of the Laplace angle within $2.5^\circ$. These small amplitudes are characteristic of the \po state. Another key feature of these solutions is that the resonance forcing induces considerable variation in the eccentricities $e_i$ and resonance offsets $\Delta_i$ during the orbits.

The masses obtained for the three planets in the Laplace resonance ($m_\mathrm{c}=267\,m_\oplus$, $m_\mathrm{b}=855\,m_\oplus$, $m_\mathrm{e}=17\,m_\oplus$) and the system inclination are consistent with the studies of \citet{MillhollandEtal2018} and \citet{TrifonovEtal2018}.\footnote{All models fit the RV amplitudes. It is therefore the quantity $m_p \sin i/m_\star^{2/3}$ that must be compared among studies.} The RV amplitudes and residuals corresponding to the best fit are plotted in \fg{gj876rv}. Upon inspection of the residuals, a Lomb-Scargle analysis shows a clear signal at $P_\mathrm{d}=1.9378\,\mathrm{d}$ with amplitude $K_\mathrm{d}=6.000\,\mathrm{m\,s^{-1}}$, revealing the presence of the inner planet d. Accounting for planet d further reduces the residuals of our fit by a factor of 2, rendering our \fpo-constrained $\chi^2$ on par with the best-fit models presented by \citet{MillhollandEtal2018}.

\citet{MillhollandEtal2018} already surmised that the \gj Laplace resonance may coincide with a fixed point in the resonance phase space. 
Measurement noise can shift an unconstrained best fit away from a fixed point \citep{JensenMillholland2022} by fitting \fpos directly, we can still test whether the data are consistent with that state.

\subsection{HR 8799}
\begin{table*}
	\centering
	\caption{Selected grid search results for HR 8799.}
	\label{tab:hr8799cid}
	\begin{tabular}{lrrrrr||rrrrr|rr}
id & $\Delta_1^\dagger$ & ${m}_\mathrm{e}$ [$m_\star$] & ${m}_\mathrm{d}$ [$m_\star$] & ${m}_\mathrm{c}$ [$m_\star$] & ${m}_\mathrm{b}$ [$m_\star$] & $a_\mathrm{e}$ [mas] & $\omega_\mathrm{e}$ [$^\circ$] & $i$ [$^\circ$] & $\Omega$ [$^\circ$] & $\tilde{G}$ & $\chi^2$ & $\chi^2_\mathrm{optm}$ \\
\hline
$ 1$ & $0.012$ & $0.00316$ & $0.00562$ & $0.00562$ & $0.00100$ & $395.49$ & $153.3$ & $155.6$ & $33.8$ & $6.407$ & $1840$ & $1150$ \\
$ 2$ & $0.011$ & $0.00316$ & $0.00562$ & $0.00562$ & $0.00316$ & $396.58$ & $146.2$ & $153.1$ & $31.2$ & $6.480$ & $2410$ & $\rightarrow$1 \\
$ 3$ & $9.4{\times}10^{-3}$ & $0.00316$ & $0.00562$ & $0.00562$ & $0.00316$ & $396.74$ & $141.8$ & $152.8$ & $27.1$ & $6.915$ & $2670$ & $\rightarrow$1 \\
$ 4$ & $8.0{\times}10^{-3}$ & $0.00316$ & $0.00562$ & $0.00562$ & $0.00562$ & $398.08$ & $135.8$ & $150.1$ & $26.0$ & $6.967$ & $2810$ & $\rightarrow$1 \\
$ 5$ & $5.2{\times}10^{-3}$ & $0.0100$ & $0.00562$ & $0.00100$ & $0.00316$ & $395.44$ & $140.5$ & $153.0$ & $25.1$ & $6.851$ & $3390$ & $\rightarrow$1 \\
$ 6$ & $0.011$ & $0.00316$ & $0.00562$ & $0.00562$ & $0.00178$ & $395.41$ & $146.6$ & $154.7$ & $28.5$ & $6.766$ & $3650$ & $\rightarrow$1 \\
$ 7$ & $9.4{\times}10^{-3}$ & $0.00562$ & $0.00562$ & $0.00562$ & $0.0100$ & $391.80$ & $138.2$ & $151.0$ & $20.9$ & $6.748$ & $4030$ & $\rightarrow$4 \\
$ 8$ & $8.0{\times}10^{-3}$ & $0.00562$ & $0.00562$ & $0.00562$ & $0.0100$ & $393.06$ & $136.0$ & $150.2$ & $19.6$ & $7.078$ & $4260$ & $\rightarrow$5 \\
$ 9$ & $8.0{\times}10^{-3}$ & $0.00316$ & $0.00316$ & $0.00316$ & $0.00562$ & $403.88$ & $129.4$ & $149.2$ & $26.4$ & $7.040$ & $4470$ & $\rightarrow$7 \\
$10$ & $9.4{\times}10^{-3}$ & $0.00178$ & $0.00562$ & $0.00562$ & $0.00178$ & $400.70$ & $143.2$ & $152.2$ & $32.3$ & $6.808$ & $4490$ & $\rightarrow$1 \\
\hline
$ 1$ & $0.010$ & $0.00347$ & $0.00586$ & $0.00581$ & $0.00328$ & $395.14$ & $145.3$ & $153.4$ & $28.7$ & $6.614$ & $1150$ \\
\hline
\end{tabular}

\end{table*}
\begin{figure*}
	\includegraphics[width=\textwidth]{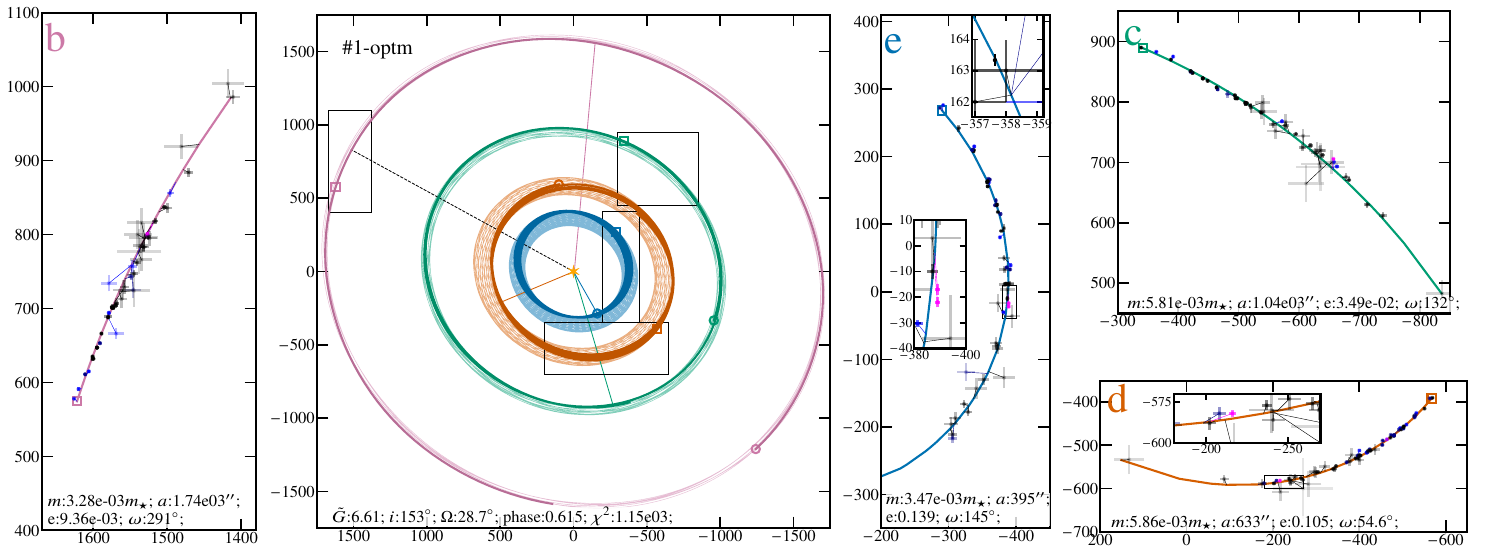}
	\caption{\label{fig:hr8799}Best-fit periodic solution for the planets in the HR 8799 system. In panels (b)--(e), the color of the error bar indicates the fit quality: ${<}2\sigma$ (black), $2{-}3\sigma$ (dark blue), $3{-}5\sigma$ (blue), $5{-}7\,\sigma$ (purple), ${>}7\sigma$ (pink). See \fg{pds70} for the meaning of the symbols in the system orbit panel. The top inset in (e) contains the GRAVITY data point.}
\end{figure*}

The \hr system is a widely spaced, iconic multi-planet system featuring four planets with orbital periods close to an 8:4:2:1 ratio.   
Because the planets move slowly, the available astrometry covers only short orbital arcs and cannot establish resonance by showing that the resonant angles librate.  
The \hr system has previously been studied using the \po approach \citep{GozdziewskiMigaszewski2020,ZurloEtal2022}. It is therefore instructive to see whether our independent modelling approach finds the same solution. Motion in the \hr system is anti-clockwise on the sky plane, which in our setup corresponds to inclinations ${>}90^\circ$ (\Tb{skypars}). We adopt broad ranges for the sky parameters (see \Tb{skypars}). The adopted range for $\tilde{G}$ extends well beyond the uncertainties in the distance to \hr and estimates for its stellar mass \citep{KervellaEtal2022}.

Results are presented in \Tb{hr8799cid}. Similar to \gj, the on-grid $\chi^2$ values span a wide range: neighboring grid points can easily be separated by ${\sim}1\,000$ in $\chi^2$. Allowing the planet masses and $\Delta_1^\dagger$ to vary continuously in the subsequent optimization reveals a global minimum at $\chi^2=1150$.
In fact, each of the ten nodes listed in \Tb{hr8799cid} converges to this minimum, signifying that the $\chi^2$ parameter space is characterized by deep valleys with the grid points corresponding to locations on the ridges. The orbit corresponding to the $\chi^2$ minimum is shown in \fg{hr8799}.

The obtained $\chi^2=1150$ is substantially lower than the $\chi^2$ reported for any of the solutions presented by \citet{ZurloEtal2022}, despite using the same astrometric data. This discrepancy likely arises from differences in the adopted astrometric uncertainties (see \app{astromdata}). Evaluating their solutions using our adopted uncertainties yields $\chi^2=1270$, bringing the results into much closer agreement. Our best-fit \po solution nevertheless implies slightly different planet masses. Assuming $m_\star=1.5\,m_\odot$, we obtain $(m_\mathrm{e},m_\mathrm{d},m_\mathrm{c},m_\mathrm{b})=(5.4,9.2,9.1,5.2)\,m_\mathrm{J}$. These values are approximately $2\,m_\mathrm{J}$ lower for planet e and $1\,m_\mathrm{J}$ higher for planet c than those reported by \citet{ZurloEtal2022}. 
We have verified, however, that the underlying \po configuration (panel (f) of \fg{fpos}) is essentially the same as that of \citet{GozdziewskiMigaszewski2020}, on which the analysis of \citet{ZurloEtal2022} is based. We therefore are consistent with their results, with the remaining differences likely arising from the adopted uncertainties in the astrometric data.

The reduced $\chi^2_\nu = 2.7$ formally indicates that the model does not provide a satisfactory fit. It will therefore be worthwhile to examine how \po solutions evolve as new astrometric data become available, and to quantify the improvement obtained by relaxing the \po constraint. Recently, an inner super-Jupiter companion, planet f ($m_\mathrm{f} \sim10{-}14\,m_\mathrm{J}$ at ${\sim}2{-}3\,\mathrm{au}$), has been proposed on the basis of Gaia stellar absolute astrometry, with additional constraints from RV and direct-imaging detection limits \citep{LagrangeEtal2026}. The inner companion is dynamically decoupled from the four outer planets. The corresponding stellar reflex motion would introduce a signal of order $0.5\,\mathrm{mas}$ into the relative astrometry of all four outer planets, but this is smaller than the uncertainties of all but the single GRAVITY measurement. We therefore expect its effect on most of the data to be small. Nevertheless, extending our model to incorporate epoch astrometry from Gaia DR4 would be a worthwhile follow-up, enabling a more complete description of the system.

\section{Discussion}
In this work, we have introduced the \fpo model as a tool for fitting resonant multi-planet systems and demonstrated its utility in constraining planetary orbits and masses. We applied the method to the \pds, \gj, and \hr systems, obtaining fits through $\chi^2$ minimization that adequately reproduce the data. This success relied on constructing $(n_p{+}1)$-dimensional data cubes containing \fpo data. For two planets, computing the \fpos is straightforward. As the number of planets in the resonant chain increases, however, the dimensionality of the problem makes this approach increasingly demanding.  
For the four-planet \hr system, we used only seven mass-grid values, starting at ${\sim}1$ Jupiter mass (see \Tb{datacubes}); nevertheless, all ten on-grid solutions listed in \Tb{hr8799cid} converged to the same off-grid $\chi^2$ minimum. For \gj, by contrast, only about half of the on-grid trials converged to the best-fit minimum. A coarser grid could miss the relevant basin entirely, underscoring the need for sufficiently dense sampling. 
Chains containing four (perhaps five) or more planets probably favor a targeted approach, rather than agnostic grid searches.

A more fundamental limitation of the \fpo method is the assumption that the planets in a resonant chain precess apsidally at a common rate. 
This need not hold in more complex systems. 
For TRAPPIST-1, \citet{AgolEtal2021} and \citet{TeyssandierEtal2022} found two apsidal precession periods in their best-fitting TTV solutions---approximately 0.032 yr for planets b and c and 1.3 yr for the outer five---contrary to our assumption of a common precession rate.
Kepler-221 provides another example: three planets form a 6:3:1 resonant chain, while an intermediate planet does not participate in the resonance.  
Configurations consistent with this state \citep{YiEtal2025} indicate that the inner pair, in the 2:1 resonance, precesses at a common rate, whereas the outer pair, in the 3:1 resonance, does not.  
In K2-138, a first-order three-body angle, $\phi_i=\sum_j k_j\lambda_j-\varpi_i$ with $\bm{k}=(2,-4,3)$, librates while the Laplace angle does not \citep{CerioniBeauge2023}. 
If an \fpo with common apsidal precession exists, its winding vector must satisfy $\bm{N}\cdot\bm{k}=0$ and $\bm{N}\cdot\bm{l}\neq0$, where $\bm{l}=(4,-7,3)$ describes the Laplace angle. 
At the same time, $N_1/N_2$ and $N_2/N_3$ must approximate the observed period ratios $P_2/P_1$ and $P_3/P_2$. 
Satisfying these requirements may demand impractically large winding numbers.

These examples suggest that complex resonant chains may require more flexible models. When the planets do not share a common apsidal precession rate, one possibility is to fit dynamically coherent subsets separately. In TRAPPIST-1, this would mean seeking an \fpo solution for planets b and c and another for the outer five, each with its own $t_\mathrm{po}$, then testing whether the coupled subsets preserve the observed resonances.

Nevertheless, for systems of moderate complexity near first-order integer commensurabilities, the \fpo method offers an attractive approach to fitting planetary orbits for both physical and practical reasons. Physically, as stated in \se{amd}, we expect planets to settle into such states after experiencing sufficient damping. For young, distant planets, agreement with an \fpo would therefore support the view that damping during planet formation led to resonant trapping. For evolved, close-in planets, \fpos could instead indirectly constrain the efficiency of tidal damping.

The \fpo method also offers substantial practical advantages. First, it is designed to yield stable solutions. Second, it substantially reduces the number of fitting parameters. For the second planet, only one orbital degree of freedom remains: the semi-major axis (here parameterized by $\Delta_1^\dagger$); for the third planet and beyond, all four coplanar orbital elements are determined by the \fpo solution. This reduction greatly simplifies the parameter search. Finally, because \fpo orbits repeat after $t_\mathrm{po}$, an integration spanning one period suffices to predict their later evolution (see \se{gj876}). This is especially useful for close-in transiting planets and RV systems observed over many orbital periods.

We then applied the \fpo method to three iconic exoplanet systems: \hr, \pds, and \gj. For \hr, we independently recovered a previously known solution \citep{ZurloEtal2022}, demonstrating the reliability of our fitting strategy. Although the available astrometric data cover only partial orbital arcs, a single, well-defined $\chi^2$ minimum emerges, illustrating the effectiveness of the \fpo method. Nevertheless, the best-fitting solution has a reduced $\chi^2_\nu = 2.7$, well above unity. It would therefore be worthwhile to revisit this system as further data become available and to investigate whether relaxing the strict \fpo assumption would improve the fit.

For \pds, the available astrometric data do not identify a unique solution: many \fpos spanning a range of masses and eccentricities fit the observations. Nevertheless, these solutions fall into two groups: Group I, in which the inner planet is more massive, and Group II, in which the outer planet is more massive. The two groups also differ in their inferred values of the longitude of the ascending node and the stellar mass. Although Group I yields slightly better fits to the astrometry, independent constraints on $\Omega$ and $m_\star$ favor Group II, with the more massive outer planet. The \fpo method thus already provides testable predictions for the planet masses and orbits. Further astrometric observations could help distinguish between these two solution families. 
 If the recently suggested third planet \citep{Balsalobre-RuzaEtal2026} can be incorporated into the \fpo model, it could further constrain the system's orbital parameters.

For \gj, applying the \fpo model to the RV data compiled by \citet{MillhollandEtal2018} yields perhaps the most striking result. The observed RV signal exhibits variations on two distinct timescales: a short timescale comparable to the orbital periods and a much longer timescale of ${\sim}8\,\mathrm{yr}$. We have shown that both arise naturally from the best-fitting \fpo model: the short timescale reflects the period of the \fpo ($t_\mathrm{po}$), while the long timescale corresponds to its precession period $2\pi t_\mathrm{po}/\omega_\mathrm{prec}$. Although a direct comparison with the fit of \citet{MillhollandEtal2018} is not possible without implementing their stellar noise model, the \fpo solution appears to reproduce the observations similarly well and readily reveals the signal of the inner non-resonant planet d. This agreement is particularly striking given how restrictive the model is. A conventional search with broadly uniform priors would be unlikely to recover this particular solution, as it occupies only a small volume of parameter space among many similarly good, non-\fpo fits. Turning the argument around, the success of the \fpo model suggests that \gj lies at or near the fixed state, with its Laplace resonance angle librating at low amplitude. This interpretation implies that the Laplace resonance is not chaotic, despite earlier evidence to the contrary \citep{BatyginEtal2015}.

Naturally, the \fpo model could also be applied to other exoplanet detection methods, such as transit timing variations (TTVs) and absolute astrometry. TTVs have mostly been analyzed with orbit-averaging approaches \citep{LithwickEtal2012,NesvornyVokrouhlicky2016}. 
However, orbit-averaged approaches do not capture short-period chopping signals in TTVs, which could be modeled within the \fpo framework (Huang et al., in prep.).   
Turning to absolute astrometry, the fourth Gaia data release could offer new opportunities to identify resonant planets on orbits of order 1 au.
A 2:1 resonant pair can produce an astrometric signal similar to that of a single planet \citep{GreenfieldOwen2026}.
\Fpos may help distinguish the two configurations by constraining the allowed orbits.

Finally, the most obvious extension of the resonance model is to relax the assumption that planets are in the \fpo state.  After all, when eccentricity damping is insufficient, resonantly trapped planets will exhibit librations in their orbital parameters (\fg{amd}) on timescales ${\gg}t_\mathrm{po}$. In this picture of the resonant landscape \fpos form special points around which resonant configurations librate. A more inclusive model for fitting resonant systems would allow configurations close to an \fpo rather than require strict conformity to the \fpo state. We will pursue this in subsequent work (Paper II; Yi et al.~in prep.).

\section{Conclusions}
Fixed-point (or periodic) orbits are resonant solutions in which the relative configuration of the planets repeats after a finite interval, modulo an overall rotation. This fixed-point property allows us to determine these orbits numerically, linking the planet masses and orbital spacings within the resonance chain to the orbital configuration. Each \fpo solution also specifies the duration of the \fpo cycle and the precession rate.

We populated multidimensional grids with these \fpo solutions and used them to fit archival astrometric and radial-velocity data for planet systems discovered through direct imaging and radial-velocity measurements. Our main conclusions are:

\begin{enumerate}
	\item Resonant planet systems that experience sufficient eccentricity damping, for example during disk-driven migration or subsequent tidal evolution, approach \fpo states. If these configurations undergo little subsequent disruption, they may remain close to these states, motivating the use of \fpo models to fit their observed orbits.

	\item The \fpo fitting approach greatly reduces the number of free parameters and can substantially reduce the computational cost. Integrating a solution over a single \fpo cycle of duration $t_\mathrm{po}$ suffices to reconstruct the state at any other time. When the observations span timescales ${\gg}t_\mathrm{po}$, as in the \gj system, using the \fpo model becomes particularly advantageous.

	\item The \fpo solutions generally agree with previously described relations among planet masses, eccentricities, and resonance offsets. For the 2:1 resonance at small $\Delta$, we identify a configuration in which conjunction occurs when both planets are at periapsis. We also find that additional, as yet undetected outer planets in the resonance chain can increase the eccentricities of the inner planets.

	\item For the \pds system the two-planet fit fall in two groups. In Group I the inner planet b is inferred to be more massive than c and the $\chi^2$ is the lowest. Conversely, in the second group the mass of planet c is on average higher than that of planet b. Group II is favored, because its inferred stellar mass and nodal angle $\Omega$ are in line with values reported in other studies.

	\item For \gj our \fpo solution fits existing RV data to levels similar as previous studies. This strongly suggest that the \gj 4:2:1 Laplace resonance is in (or close) to the fixed point and its resonance is not chaotic.

	\item For \gj the RV-residuals after fitting the three-planet resonance chain readily point to the presence of the dynamically detached inner planet d, illustrating that the \fpo method could also identify additional perturbers.

	\item For \hr, we recover a previously reported \fpo configuration. Epoch astrometry from the forthcoming Gaia DR4 will enable a joint fit to relative and absolute astrometry, offering further constraints on the system.
\end{enumerate}
Even when a planet system is not in an \fpo state, a stable \fpo solution can provide a useful starting point for more refined modelling of its resonant configuration. In Paper II, we will explore this approach using a metric that quantifies the distance between an orbital fit and an \fpo solution.

Software: REBOUND: \citet{ReinLiu2012}, REBOUNDx: \citet{TamayoEtal2020}, NUMPY: \citep{VanDerWaltEtal2011}, MATPLOTLIB: \citet{Hunter2007}, PANDAS: \citet{McKinney2010}, ASTROPY: \citep{AstropyCollaborationEtal2022} 

\begin{acknowledgements}
This work was supported by the National Natural Science Foundation of China (NSFC) project no. 12473065 and 12233004. The authors greatly appreciate discussion and input from  Valentin Christiaens, Iain Hammond, Yukun Huang, Haochang Jiang, Michiel Lambrechts, Adrien Leleu, Rosemary Mardling, Hanno Rein, Lauren Weiss, and Alan Cefali Signor. S. H. thanks K.C. Wong Education Foundation for supporting TDLI postdoc research. 
\end{acknowledgements}

\bibliographystyle{aa}

\bibliography{aa}

\begin{appendix}

\section{Finding the \po solutions} \label{app:po-finder}
\begin{figure*}
	\sidecaption
	\includegraphics[width=0.7\columnwidth]{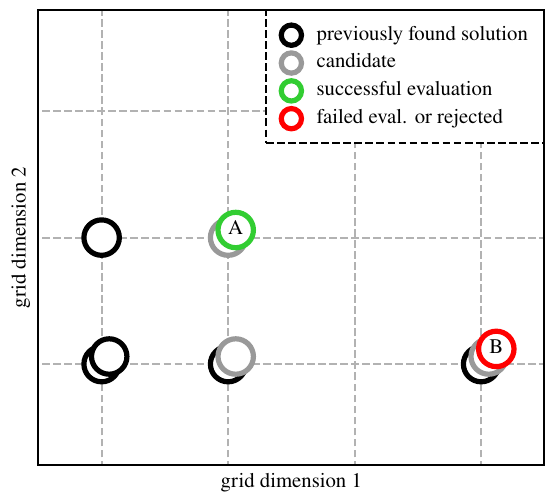}
	\includegraphics[width=0.7\columnwidth]{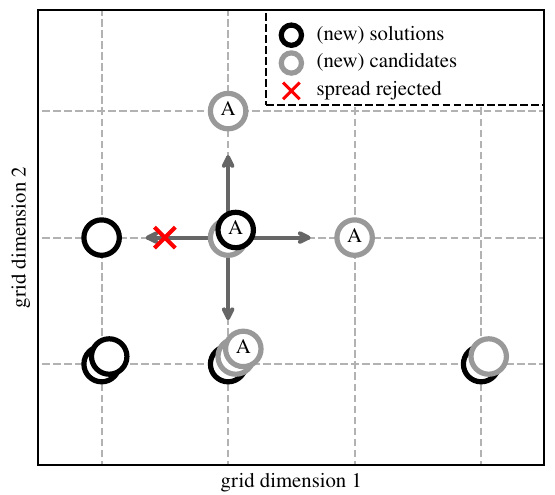}
	\caption{\label{fig:stones}Strategy for extending PO solutions within the data cube. \textit{Left}: the data cube consists of already vetted solutions $\bm{x}_\mathrm{sol}$ (black) and candidate solutions (grey). At a grid point, multiple solutions and candidates can coexist. Two candidates are selected for evaluation; one succeeds (green) and one fails (red). \textit{Right}: The successful solution (A) spreads as a candidate to all neighboring grid points, provided a similar (candidate) solution is not present. The rejected (candidate) solution (B) has been removed from the stack.}
\end{figure*}

\begin{table*}
	\centering
	\caption{\label{tab:datacubes}List of \fpo cubes containing the resonance configurations.}
	\begin{tabular}{lllccccc}
\hline
\noalign{\vskip 4pt}
Resonance chain & Table & $\mathbf{k}$ & $N_{q}$ & $q$ range & $N_{\Delta}$ & $\Delta$ range & $e_{\rm lim}$ \\
\noalign{\vskip 3pt}
\hline
2:1 & \texttt{21low} & $(-1,2)$ & 15 & $[10^{-5},\,10^{-1.5}]$ & 34 & $[-0.258,\,0.341]$ & 0.9 \\
2:1 & \texttt{21med} & $(-1,2)$ & 22 & $[10^{-5},\,10^{-1.5}]$ & 50 & $[-0.212,\,0.386]$ & 0.5 \\
3:2 & \texttt{32low} & $(-2,3)$ & 15 & $[10^{-5},\,10^{-1.5}]$ & 33 & $[-0.258,\,0.258]$ & 0.9 \\
3:2 & \texttt{32med} & $(-2,3)$ & 22 & $[10^{-5},\,10^{-1.5}]$ & 49 & $[-0.212,\,0.316]$ & 0.5 \\[5pt]
3:2:1 & \texttt{321low} & \begin{tabular}[t]{@{}l@{}}$(1,-2,1)$ \\ $(-1,1,1)$ \\ $(1,-3,3)$\end{tabular} & 15 & $[10^{-5},\,10^{-1.5}]$ & 33 & $[-0.258,\,0.258]$ & 0.5 \\[5pt]
4:2:1 & \texttt{421low} & \begin{tabular}[t]{@{}l@{}}$(1,-3,2)$ \\ $(-1,1,2)$ \\ $(-2,5,-2)$ \\ $(1,-4,4)$\end{tabular} & 15 & $[10^{-5},\,10^{-1.5}]$ & 34 & $[-0.258,\,0.341]$ & 0.5 \\[5pt]
8:4:2:1 & \texttt{8421low} & \begin{tabular}[t]{@{}l@{}}$(1,-2,-1,2)$ \\ $(-1,1,3,-2)$ \\ $(1,-3,1,2)$ \\ $(-1,2,-1,2)$\end{tabular} & 7 & $[10^{-3},\,10^{-1.5}]$ & 11 & $[0.001,\,0.015]$ & 0.3 \\
\hline
\end{tabular}

    \tablefoot{The planet-to-star mass ratio is defined as $q=m/M_\star$. Each planet uses the same logarithmically spaced mass-ratio grid with $N_q$ values per mass dimension, while the $\Delta$ dimension contains $N_{\Delta}$ values. The adopted eccentricity cutoff is $e_{\rm lim}$. The resonance coefficient vector $\bm{k}$ is defined in \eq{k-vector}. The cubes are available online at \url{https://github.com/chrisormel/periodic-orbits}. }
\end{table*}

To find the roots to \eq{po-eq} a Newton-Raphson method is adopted \citep{PressEtal2007}, such that around a guess solution $\bm{x}_c$ \eq{po-eq} is linearized
\begin{equation} \label{eq:linearized}
	\bm{F}_N(\bm{x}_c) +\left(\frac{\partial \bm{F}_N}{\partial \bm{x}}\right)_{\bm{x}_c} (\bm{x} -\bm{x}_c) -\bm{x} = 0,
\end{equation}
where $\bm{F}_N(\bm{x})$ represents integration of $\bm{x}$ over $N$ periapse passages of the inner-most body and $\partial \bm{F}/\partial{\bm{x}}$ is the Jacobian of this mapping. \Eq{linearized} provides a new estimate to the solution $\bm{x}_\mathrm{sol}$. Convergence is usually rapid when the candidate solution $\bm{x}_c$ lies sufficiently close to the true solution $\bm{x}_\mathrm{sol}$ (Jacobian evaluations can often be reused). Conversely, when the candidate is far from the root, the Newton-Raphson method will typically fail, although it occasionally happens that the algorithm "stumbles" upon a solution by chance---it converge to a root far from the initial guess.

We aim to obtain a complete census of the solution space, also allowing for multiple independent solutions at the same grid point. The adopted strategy is illustrated in \fg{stones}. Starting from at least one known solution, we populate the grid by using the known solutions as candidate solutions at neighboring grid points. Typically, an initial solution can be sought by assuming a special geometry, e.g., an alignment of planets at $t=0$ along a single line (the so-called symmetric \fpo; \citealt{AntoniadouVoyatzis2022}).

\Fg{stones} illustrates how solutions spread across the parameter grid. At each grid point, multiple solutions ($\bm{x}_\mathrm{sol}$ satisfying \eq{po-eq}) and multiple candidates may coexist. During each cycle, one or more candidates are selected for evaluation. Employing multiprocessing techniques, multiple candidates can be processed simultaneously, provided that they are separated by at least one grid point. The Newton-Raphson iteration either converges upon a solution or fails to find a root. In certain cases a solution may nevertheless be rejected (e.g., because it is long-term unstable; see below). If no root is found, or if the solution is rejected, the candidate will be removed from the stack. Otherwise, the solution is accepted and spreads to its neighboring grid nodes as new candidates (see \fg{stones}). The procedure then enters a new cycle.

\begin{figure}
	\centering
	\includegraphics[width=0.9\columnwidth]{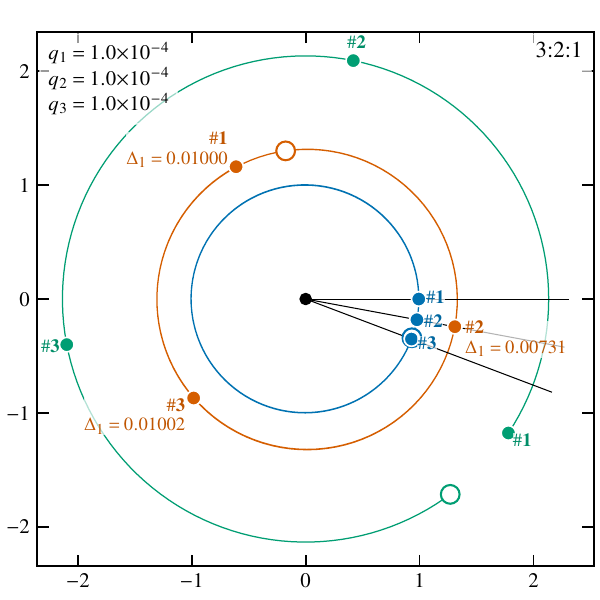}
	\caption{\label{fig:duplicates}Duplicate \fpo solutions. In this example of a 3:2:1 resonance, state \#1 represenets a proper \fpo solution: if integrated over 3 periapse passages, over a time $t=t_\mathrm{po}$, it returns to its intitial state (open circles) modulo a rotation. During this cycle, the inner-most orbit is also at periapses at two intermediate times (\#2 and \#3). Hence, there are in total 3 equivalent solutions for the same \fpo. We only accept one.}
\end{figure}
Propagating solutions in this way can result in a rapid growth in the number of candidates, particularly in higher-dimensional grids corresponding to a larger number of planets.  To avoid redundant searches, screen for stability, and impose physical restrictions, the following rejection conditions are applied:
\begin{itemize}
	\item The solution already exists in the stack. For example, if solution A in \fg{stones} satisfies $||\bm{x}_A - \bm{x}_i|| < \epsilon$ for any existing solution $\bm{x}_i$ at the same grid point, we discard it. Furthermore, since $\bm{x}_i$ has already been propagated, there is also no reason for spreading A. Likewise, if a candidate lies within $\epsilon$ of the newly found A, we remove the candidate. A similar uniqueness criterion is also applied when the solution is added to the candidate list of a neighboring grid point. Here we choose $\epsilon=0.01$.
	\item The relevant resonance angles do not librate. It is possible, though unlikely, that the \fpo solution is non-resonant or is associated with another resonance than the one under investigation. We therefore verify whether the resonances associated with the resonance chain actually librate over the span of one periodic orbit.
    We consider the resonance condition satisfied if at least one associated resonance angle librates, defined as remaining confined to an angular interval narrower than $0.9\times2\pi$.
       For two planets, the joint resonance angle can librate, while the two classical angles do not (see \se{sol-cubes}).
	\item The \fpo solution is unstable. An \fpo is not guaranteed to be long-term stable and its stability must be assessed by perturbing the solution and integrating it on timescales ${\gg}t_\mathrm{po}$.  Because integrating every perturbed solution over the same long duration would be computationally expensive, we use here a streamlined procedure to identify likely stable solutions.
	      
	      For each stability trial, we initialize the system at $\bm{x} = \bm{x}_\mathrm{fPO} + \delta \bm{x}_\mathrm{fPO}$ and integrate it for $t_\mathrm{X}=10^3\,t_\mathrm{po}$. If necessary, we adjust the endpoint of the integration to ensure that the innermost planet returns exactly to periapsis to enable comparison with the original \fpo (comparison proceeds in terms of the $\bm{x}_\mathrm{sol}$ parameters listed in \Tb{coordinates}).  We consider the \fpo stable if both of the following conditions hold:
	      \begin{enumerate}
		      \item[(i)] The deviation from the original \fpo satisfies $\epsilon=|\bm{x}-\bm{x}_\mathrm{fPO}|\leq\epsilon_\mathrm{stab}$, where $\epsilon_\mathrm{stab}=0.02$ has been adopted; and
		      \item[(ii)] The running maximum of the deviation $\epsilon_\mathrm{max}$ has not increased significantly with respect to $t=t_X/2$, $\epsilon_\mathrm{max}(t_\mathrm{X}) \le 1.01\epsilon_\mathrm{max}(t_\mathrm{X}/2)$.
	      \end{enumerate}
	      If condition (i) is violated at any point, we reject the solution as unstable. Otherwise, if condition (ii) has not been met, we double $t_\mathrm{X}$, and continue the integration, repeating this procedure for up to four doublings. If condition (i) remains satisfied after four doublings but condition (ii) is still not met, the test has not established whether the deviation remains bounded. In practice, these cases often exhibit slow, approximately linear growth, with deviations remaining several orders of magnitude below $\epsilon_\mathrm{stab}$ throughout the integration. Such approximately linear growth is compatible with regular motion, by analogy with the linear growth of infinitesimal deviations underlying the MEGNO \citep{CincottaSam2000} diagnostic. For these reasons, we retain these solutions in our table. But it should be recognized that much more rapid growth at later times, while unlikely, cannot be excluded (see \se{sol-cubes}).

	      For each \fpo, we repeat this procedure with seven randomly selected initial perturbations $\delta \bm{x}_\mathrm{fPO}$, because the growth of $\epsilon$ depends strongly on the initial displacement. We reject the \fpo as unstable if any trial violates condition (i). Stability calculations are conducted with the \texttt{whfast} and \texttt{whfast512} integrators \citep{ReinTamayo2015, Javaheri2023, DagliRein2026}.

	\item A duplicate solution exists. During one \fpo cycle, the innermost planet passes through periapsis $n$ times; choosing any of these passages as $t=0$ gives a different representation of the same \fpo, with the other planets at different orbital phases (see \fg{duplicates}). Including all such copies would unnecessarily inflate the solution cubes. Only the \fpo with the smallest $|\theta_n -\theta_1|$ (i.e., the difference in true longitude between the inner- and outer-most planet\footnote{We take care of the branch cut at 0, such that $\theta_n=6.0$ and $\theta_1=0.0$, for example, will give a difference of $0.28$, not $6.0$.}) is included in our table. In \fg{duplicates} this corresponds to configuration \#1.
	\item Physical restriction. In particular, we impose a maximum eccentricity $e_\mathrm{lim}$ and discard solutions for which the eccentricity of one or more bodies exceeds $e_\mathrm{lim}$. At high eccentricities, the solution space becomes strongly multimodal, so these configurations can dominate the search and the resulting solution cubes. These high-eccentricity configurations, which often exhibit crossing orbits, are outside the primary focus of this work.

	      	      \end{itemize}

\begin{table}
	\centering
	\caption{\label{tab:Laplace}Laplace coefficients.}
    \begin{tabular}{lcc}
    \hline
    Resonance & $r_1$ & $r_2$ \\
    \hline
    2:1   &$-1.190$ & $0.428$ \\
    3:2   &$-2.025$ & $2.484$ \\
    \hline
    \end{tabular}
    \tablefoot{Adopted from \citet{TerquemPapaloizou2019}.}
\end{table}

\subsection{The solution cubes} \label{sec:sol-cubes}

\Tb{datacubes} summarizes the grids adopted for the 2:1, 3:2, 4:2:1, 3:2:1, and 8:4:2:1 resonance configurations. We have made these tables publicly available at \url{https://github.com/chrisormel/periodic-orbits}. The labels \texttt{low} and \texttt{med} denote coarse and finer grid resolutions, respectively. The eccentricity cutoff varies between cubes; solutions exceeding the corresponding $e_{\rm lim}$ for any planet are excluded as described above. The retained solutions satisfy our stability criteria. Importantly, it must be emphasize that  these criteria do not guarantee all \fpos remain long-term stable. Additional integrations extending to $10^5$ inner-planet orbital periods reveal rare cases of resonance loss or planetary ejection in fewer than $0.1\%$ of the tested solutions. We advice the users to thoroughly test the solutions they extract from the tables. Here, the main difficulty is that some \fpo solutions exhibit instability only along specific perturbation directions $\delta \bm{x}_\mathrm{fPO}$. The \fpo solutions that we find in \se{pofit} show stability for at least 20 directions and for at least $10^5$ inner orbits.

For each resonance chain, we specify a set of integer coefficient vectors $\bm{k}$ defining resonant angles of the form
\begin{equation} \label{eq:k-vector}
    \phi = \sum_i k_i \lambda_i - g(\bm{\omega}),
\end{equation}
where $\lambda_i$ is the mean longitude of planet $i$ and $g(\bm{\omega})$ is a linear combination of the longitudes of periapse. We choose the sign of $\bm{k}$ such that $o=\sum_i k_i\geq 0$. Here, $o$ is the resonance order and the coefficients in $g$ also sum to $o$.

For two-body resonances, we also consider the combined resonant angle, whose associated longitude of periapse is defined as
\begin{equation}
    \omega_\mathrm{joint} = \mathrm{arg}\left(
        r_1 e_1 \exp[\mathrm{i}\omega_1] + r_2 e_2 \exp[\mathrm{i}\omega_2]
            \right)
\end{equation}
where $r_i$ are the Laplace coefficients, which depends on the resonance (\Tb{Laplace}).

For example, the 2:1 resonance has $\bm{k}=(-1,2)$, for which we consider three angles:
\begin{align}
    \phi_1 &= -\lambda_1 + 2\lambda_2 - \omega_1, \\
    \phi_2 &= -\lambda_1 + 2\lambda_2 - \omega_2, \\
    \phi_\mathrm{joint} &= -\lambda_1 + 2\lambda_2 - \omega_\mathrm{joint}.
\end{align}
We accept a solution as belonging to the 2:1 resonance if at least one of these angles librates.

Similarly, for the $\bm{N}=(3,2,1)$ chain, we choose $\bm{k}$ that give resonances of at most first order. This includes the zeroth-order Laplace angle associated with $\bm{k}=(-1,2,-1)$,
\begin{equation}
    \phi_L = -\lambda_1 + 2\lambda_2 - \lambda_3.
\end{equation}
as well as the first-order angles associated with $\bm{k}=(-1,1,1)$ and $\bm{k}=(1,-3,3)$. These vectors satisfy $\bm{k} \cdot \bm{N} = 0$ where $\bm{N}=(3,2,1)$ are the winding numbers of the planets associated with the periodic orbit. Each of these two vectors defines three angles, obtained by subtracting $\omega_1$, $\omega_2$, or $\omega_3$, giving seven angles in total. Under our classification, an \fpo is accepted as a solution to the 3:2:1 chain if at least one of these seven angles librates. Libration of one three-body angle does not necessarily imply that both adjacent pairs are in two-body resonance.

\section{Astrometric and RV Data} \label{app:astromdata}
\begin{figure}
	\centering
	\includegraphics[width=\columnwidth]{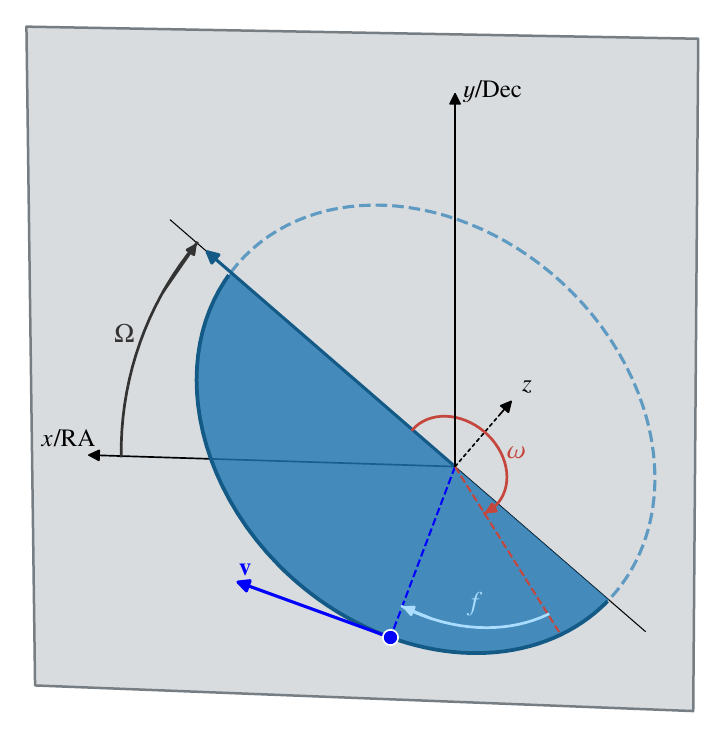}
    \caption{\label{fig:coordsystem}Coordinate system adopted in this work. We use a right-handed system with the $x$-axis pointing toward increasing right ascension and the $z$-axis pointing away from the observer. The longitude of the ascending node, $\Omega$, is measured in the sky plane from the positive $x$-axis to the node line, where the planet crosses toward positive $z$. The orbit shown has an inclination of $i=50^\circ$ and appears to move clockwise on the sky (orbits with $i>90^\circ$ would appear counterclockwise). The true longitude is defined as $\theta=\Omega+\omega+f$ for $i<90^\circ$ and $\theta=\Omega-\omega-f$ for $i>90^\circ$.}
\end{figure}

\subsection{Coordinate system convention}
Our coordinate convention is illustrated in \fg{coordsystem}. We adopt a right-handed Cartesian system $(x,y,z)$, with the positive $x$- and $y$-axes pointing toward increasing right ascension (RA) and declination (Dec), respectively. The $z$-axis therefore points away from the observer, so positive line-of-sight velocities indicate recession. The longitude of the ascending node, $\Omega$, is measured from the positive $x$-axis toward the positive $y$-axis, i.e., from East toward North. This differs from the usual astrometric convention, in which the position angle (PA) of the ascending node is measured from North through East.

Under our convention, orbital motion appears clockwise on the sky for $0 \le i < 90^\circ$ and counterclockwise for $90^\circ < i \le 180^\circ$.

\subsection{\pds}
\begin{table}[t]
	\centering
	\caption{Astrometric data for PDS70}
	\label{tab:pds70table}
	\small
	\begin{tabular}{rrrrl}
\hline
MJD & planet & RA (mas) & Dec (mas) \\
\hline
56017 & b & $58.7 \pm 10.7$ & $-182.7 \pm 22.2$ &  \\
56787 & b & $70.3 \pm 9.6$ & $-180.3 \pm 5.9$ &  \\
57145 & b & $83.1 \pm 3.9$ & $-173.5 \pm 4.3$ &  \\
57145 & b & $83.9 \pm 3.6$ & $-178.5 \pm 4.0$ &  \\
57174 & b & $89.4 \pm 6.0$ & $-178.3 \pm 7.1$ &  \\
57174 & b & $86.9 \pm 6.2$ & $-173.0 \pm 6.4$ &  \\
57522 & b & $90.2 \pm 7.3$ & $-170.8 \pm 8.6$ &  \\
57522 & b & $95.2 \pm 4.8$ & $-175.0 \pm 7.7$ &  \\
57540 & b & $94.5 \pm 22.0$ & $-164.4 \pm 27.6$ &  \\
58174 & b & $109.6 \pm 7.9$ & $-157.7 \pm 7.9$ &  \\
58174 & b & $110.0 \pm 7.9$ & $-157.6 \pm 8.0$ &  \\
58289 & b & $96.8 \pm 25.4$ & $-147.9 \pm 25.9$ &  \\
58642 & b & $110.9 \pm 6.8$ & $-136.4 \pm 6.8$ &  \\
58680 & b & $102.626 \pm 0.087$ & $-139.879 \pm 0.291$ &  \\
58889 & b & $104.545 \pm 0.229$ & $-135.081 \pm 0.077$ &  \\
59410 & b & $118 \pm 11$ & $-118 \pm 11$ &  \\
59447 & b & $108.5 \pm 8.2$ & $-118.0 \pm 8.2$ &  \\
59448 & b & $110.8 \pm 5.9$ & $-125.2 \pm 5.9$ &  \\
59459 & b & $109.8 \pm 5.9$ & $-128.5 \pm 6.0$ &  \\
59461 & b & $110.1 \pm 5.9$ & $-128.0 \pm 6.0$ &  \\
59638 & b & $114.3 \pm 5.2$ & $-104.0 \pm 5.2$ &  \\
59693 & b & $110.8 \pm 2.3$ & $-112.8 \pm 2.3$ &  \\
59999 & b & $110.7 \pm 6.7$ & $-97.9 \pm 6.3$ &  \\
60011 & b & $116.7 \pm 2.4$ & $-105.8 \pm 2.3$ &  \\
60011 & b & $128 \pm 11$ & $-104 \pm 11$ &  \\
60394 & b & $115.0 \pm 2.4$ & $-97.1 \pm 2.2$ &  \\
56787 & c & $-212.1 \pm 10.5$ & $66.2 \pm 7.3$ &  \\
58174 & c & $-205.0 \pm 13.0$ & $41.0 \pm 6.0$ &  \\
58289 & c & $-233.7 \pm 26.7$ & $28.7 \pm 25.0$ &  \\
58548 & c & $-222.0 \pm 8.0$ & $39.0 \pm 4.0$ &  \\
58642 & c & $-219.7 \pm 7.8$ & $40.3 \pm 8.0$ &  \\
58683 & c & $-214.98 \pm 0.296$ & $32.323 \pm 0.396$ &  \\
58683 & c & $-214.929 \pm 0.335$ & $32.222 \pm 0.437$ &  \\
58889 & c & $-214.333 \pm 0.144$ & $27.212 \pm 0.249$ &  \\
59410 & c & $-208.2 \pm 5.0$ & $12.4 \pm 3.6$ &  \\
59448 & c & $-203.6 \pm 8.0$ & $7.5 \pm 7.1$ &  \\
59459 & c & $-207.0 \pm 6.0$ & $11.0 \pm 7.2$ &  \\
59638 & c & $-195.6 \pm 9.0$ & $19 \pm 10$ &  \\
59999 & c & $-221.1 \pm 8.7$ & $-1.5 \pm 5.4$ &  \\
60011 & c & $-206.5 \pm 1.0$ & $0.0 \pm 1.0$ &  \\
60011 & c & $-196 \pm 23$ & $-1 \pm 23$ &  \\
60394 & c & $-206.4 \pm 3.0$ & $-7.2 \pm 1.8$ &  \\
59631.28 & b & $111.14 \pm 1.0$ & $-115.78 \pm 0.83$ &  \\
59222.34 & c & $-212.88 \pm 0.52$ & $19.32 \pm 0.24$ &  \\
59301.24 & c & $-212.88 \pm 0.08$ & $17.85 \pm 0.14$ &  \\
59302.25 & c & $-212.83 \pm 0.13$ & $17.67 \pm 0.10$ &  \\
59307.23 & c & $-212.88 \pm 0.16$ & $17.55 \pm 0.23$ &  \\
59363.10 & c & $-212.58 \pm 0.15$ & $16.27 \pm 0.15$ &  \\
59365.03 & c & $-212.32 \pm 0.13$ & $16.15 \pm 0.11$ &  \\
59631.28 & c & $-211.75 \pm 0.53$ & $10.04 \pm 0.37$ &  \\
59410.0 & b & $115.2 \pm 3.3$ & $-123.8 \pm 3.4$ &  \\
60406.0 & b & $116.3 \pm 4.8$ & $-97.5 \pm 4.7$ &  \\
60407.0 & b & $122.0 \pm 3.5$ & $-101.9 \pm 3.4$ &  \\
60874.0 & b & $114.6 \pm 4.0$ & $-80.4 \pm 3.2$ &  \\
60874.0 & c & $-220.1 \pm 23.2$ & $-26.1 \pm 11.0$ &  \\
60874.0 & b & $114.3 \pm 5.0$ & $-79.2 \pm 3.9$ &  \\
60874.0 & c & $-219.2 \pm 20.7$ & $-27.9 \pm 11.0$ &  \\
60878.0 & b & $122.8 \pm 2.4$ & $-83.0 \pm 1.8$ &  \\
60878.0 & c & $-218.2 \pm 13.3$ & $-22.6 \pm 7.0$ &  \\
60878.0 & b & $117.6 \pm 4.4$ & $-80.0 \pm 3.2$ &  \\
60878.0 & c & $-229.3 \pm 9.5$ & $-25.0 \pm 9.6$ &  \\
\hline
\end{tabular}

\end{table}

The astrometric data used here for \pds is listed in \Tb{pds70table}.

For PDS 70 the data is adopted from Table 2 and Table~B.1 of \citet{TrevascusEtal2025} (consult the latter table for references to the original data). Recent measurements from \citet{Balsalobre-RuzaEtal2026} have also been included. A covariance correction is not used in this work.

\subsection{\gj} \label{app:gj876data}
For \gj we adopt the vizier online catalogue associated with the \citet{MillhollandEtal2018} publication (\url{https://cdsarc.cds.unistra.fr/viz-bin/cat/J/AJ/155/106}). We have adjusted their stated velocities by including instrumental offsets and instrumental noise. That is:
\begin{align}
	\mathrm{RV}        & \rightarrow \mathrm{RV}_\mathrm{obsv} -\gamma_{\mathrm{instr},i}    \\
	\sigma_\mathrm{RV} & \rightarrow \sqrt{\sigma_\mathrm{obsv}^2 + \sigma_{\mathrm{jit}}^2}
\end{align}
such that we can construct the $\chi^2$ in the usual way. The values for $\gamma_\mathrm{instr}$, which vary across instruments, and $\sigma_\mathrm{jit}$ are not independently fitted, but adopted from \citet{MillhollandEtal2018}. We do \textit{not} adopt their Gaussian Processes stellar noise model nor correct the residuals for the interior planet d.

\subsection{\hr}
The astrometric data used to fit the \hr planets is compiled in \Tb{hr8799table}.

The data is taken from \citet{ZurloEtal2022}, their Table~A.1. Since their table does not list decimal precision (occasionally the error is listed as 0) we have consulted several original works for the accurate data. In particular, we propagated the errors in radius and position angle stated in \citep{DeRosaEtal2020} to RA and declination. A covariance correction has also not been applied.

\onecolumn

\newcommand{\foothrxxxx}{(a) Manually added these entries as...}
\newcommand{\commenta}{(a)}
\newcommand{\commentb}{(b)}
\small
\setlength{\tabcolsep}{3pt}
\begin{longtable}{rrrrl|rrrrl}
\caption{Astrometric data for HR 8799}\label{tab:hr8799table}\\
\hline
MJD & planet & RA (mas) & Dec (mas) & Note & MJD & planet & RA (mas) & Dec (mas) & Note \\
\hline
\endfirsthead

\hline
MJD & planet & RA (mas) & Dec (mas) & Note & MJD & planet & RA (mas) & Dec (mas) & Note \\
\hline
\endhead

51117.0 & b & $1411 \pm 9$ & $986 \pm 9$ &  & 54791.7 & d & $-217 \pm 20$ & $-608 \pm 20$ &  \\
51117.0 & b & $1418 \pm 22$ & $1004 \pm 20$ &  & 55043.7 & d & $-237 \pm 3$ & $-577 \pm 3$ &  \\
52472.1 & b & $1481 \pm 23$ & $919 \pm 17$ &  & 55087.5 & d & $-282 \pm 30$ & $-590 \pm 30$ &  \\
53199.0 & b & $1471 \pm 6$ & $884 \pm 6$ &  & 55109.4 & d & $-270 \pm 70$ & $-600 \pm 70$ &  \\
53568.1 & b & $1496 \pm 5$ & $856 \pm 5$ &  & 55113.0 & d & $-241 \pm 7$ & $-586 \pm 7$ &  \\
54312.7 & b & $1504 \pm 3$ & $837 \pm 3$ &  & 55134.9 & d & $-251 \pm 7$ & $-573 \pm 7$ &  \\
54396.6 & b & $1500 \pm 7$ & $836 \pm 7$ &  & 55138.6 & d & $-240 \pm 14$ & $-580 \pm 14$ &  \\
54656.3 & b & $1527 \pm 4$ & $799 \pm 4$ &  & 55390.4 & d & $-265 \pm 4$ & $-576 \pm 4$ &  \\
54689.3 & b & $1527 \pm 2$ & $801 \pm 2$ &  & 55397.7 & d & $-269 \pm 6$ & $-580 \pm 6$ &  \\
54725.9 & b & $1516 \pm 4$ & $818 \pm 4$ &  & 55499.9 & d & $-296 \pm 13$ & $-561 \pm 13$ &  \\
54791.7 & b & $1532 \pm 20$ & $796 \pm 20$ &  & 55762.7 & d & $-303 \pm 5$ & $-562 \pm 5$ &  \\
55043.7 & b & $1526 \pm 4$ & $797 \pm 4$ &  & 55850.3 & d & $-299 \pm 13$ & $-563 \pm 13$ &  \\
55058.3 & b & $1536 \pm 10$ & $785 \pm 10$ &  & 55875.9 & d & $-320 \pm 13$ & $-549 \pm 13$ &  \\
55087.5 & b & $1538 \pm 30$ & $777 \pm 30$ &  & 56128.3 & d & $-339 \pm 5$ & $-555 \pm 5$ &  \\
55109.4 & b & $1535 \pm 20$ & $816 \pm 20$ &  & 56227.1 & d & $-346 \pm 4$ & $-548 \pm 4$ &  \\
55113.0 & b & $1532 \pm 7$ & $783 \pm 7$ &  & 56230.8 & d & $-343 \pm 6$ & $-555 \pm 6$ &  \\
55134.9 & b & $1524 \pm 10$ & $795 \pm 10$ &  & 56581.3 & d & $-382 \pm 16$ & $-522 \pm 16$ &  \\
55138.6 & b & $1540 \pm 19$ & $800 \pm 19$ &  & 56588.6 & d & $-377 \pm 7$ & $-538 \pm 11$ &  \\
55390.4 & b & $1532 \pm 5$ & $783 \pm 5$ &  & 56614.2 & d & $-370.4 \pm 1.4$ & $-539.4 \pm 1.4$ & \commenta \\
55397.7 & b & $1547 \pm 6$ & $757 \pm 9$ &  & 56851.4 & d & $-400 \pm 4$ & $-512 \pm 4$ &  \\
55499.9 & b & $1535 \pm 15$ & $766 \pm 15$ &  & 56851.4 & d & $-391 \pm 2$ & $-530 \pm 2$ &  \\
55762.7 & b & $1541 \pm 5$ & $762 \pm 5$ &  & 56855.1 & d & $-400 \pm 11$ & $-534 \pm 11$ &  \\
55850.3 & b & $1579 \pm 11$ & $734 \pm 11$ &  & 56884.3 & d & $-396 \pm 1$ & $-524 \pm 2$ &  \\
55875.9 & b & $1546 \pm 11$ & $725 \pm 11$ &  & 56913.5 & d & $-397.1 \pm 1.9$ & $-529.9 \pm 1.8$ & \commenta \\
56128.3 & b & $1545 \pm 5$ & $747 \pm 5$ &  & 56997.5 & d & $-400 \pm 2$ & $-523 \pm 2$ &  \\
56227.1 & b & $1549 \pm 4$ & $743 \pm 4$ &  & 56997.5 & d & $-399 \pm 4$ & $-525 \pm 4$ &  \\
56230.8 & b & $1558 \pm 6$ & $729 \pm 9$ &  & 56997.5 & d & $-400 \pm 4$ & $-523 \pm 4$ &  \\
56581.3 & b & $1545 \pm 22$ & $724 \pm 22$ &  & 56997.5 & d & $-403 \pm 4$ & $-524 \pm 4$ &  \\
56588.6 & b & $1562 \pm 8$ & $713 \pm 13$ &  & 57209.1 & d & $-424 \pm 4$ & $-509 \pm 3$ &  \\
56851.4 & b & $1570 \pm 3$ & $704 \pm 3$ &  & 57209.1 & d & $-417 \pm 1$ & $-517 \pm 1$ &  \\
56855.1 & b & $1560 \pm 13$ & $725 \pm 13$ &  & 57234.7 & d & $-419 \pm 2$ & $-516 \pm 1$ &  \\
56913.5 & b & $1568.5 \pm 3.2$ & $706.9 \pm 4.3$ & \commenta & 57260.2 & d & $-436 \pm 11$ & $-510 \pm 12$ &  \\
56997.5 & b & $1575 \pm 2$ & $702 \pm 4$ &  & 57293.1 & d & $-420 \pm 4$ & $-513 \pm 4$ &  \\
56997.5 & b & $1574 \pm 3$ & $701 \pm 2$ &  & 57293.1 & d & $-426 \pm 1$ & $-512 \pm 1$ &  \\
56997.5 & b & $1574 \pm 4$ & $701 \pm 3$ &  & 57651.5 & d & $-453.1 \pm 1.8$ & $-499.2 \pm 1.7$ & \commenta \\
56997.5 & b & $1573 \pm 3$ & $701 \pm 3$ &  & 57710.1 & d & $-464 \pm 1$ & $-486 \pm 2$ &  \\
57209.1 & b & $1579 \pm 1$ & $694 \pm 1$ &  & 57710.1 & d & $-454 \pm 2$ & $-489 \pm 2$ &  \\
57234.7 & b & $1580 \pm 5$ & $689 \pm 3$ &  & 57918.3 & d & $-473 \pm 2$ & $-476 \pm 2$ &  \\
57260.2 & b & $1569 \pm 11$ & $666 \pm 7$ &  & 57918.3 & d & $-472 \pm 2$ & $-482 \pm 2$ &  \\
57293.1 & b & $1580 \pm 1$ & $688 \pm 1$ &  & 58038.7 & d & $-480 \pm 2$ & $-478 \pm 2$ &  \\
57710.1 & b & $1589 \pm 2$ & $666 \pm 1$ &  & 58038.7 & d & $-480 \pm 1$ & $-477 \pm 1$ &  \\
57918.3 & b & $1591 \pm 1$ & $653 \pm 1$ &  & 58038.7 & d & $-492 \pm 5$ & $-463 \pm 6$ &  \\
58038.7 & b & $1595 \pm 1$ & $647 \pm 1$ &  & 58038.7 & d & $-480 \pm 1$ & $-477 \pm 1$ &  \\
58038.7 & b & $1595 \pm 1$ & $647 \pm 1$ &  & 58286.9 & d & $-495 \pm 1$ & $-460 \pm 2$ &  \\
58286.9 & b & $1601 \pm 1$ & $635 \pm 1$ &  & 58286.9 & d & $-497 \pm 2$ & $-463 \pm 2$ &  \\
58349.0 & b & $1601 \pm 2$ & $632 \pm 3$ &  & 58349.0 & d & $-509 \pm 2$ & $-452 \pm 3$ &  \\
58349.0 & b & $1600 \pm 1$ & $632 \pm 1$ &  & 58349.0 & d & $-502 \pm 1$ & $-461 \pm 1$ &  \\
58786.9 & b & $1606 \pm 2$ & $615 \pm 2$ &  & 58349.0 & d & $-503 \pm 2$ & $-456 \pm 2$ &  \\
58790.6 & b & $1611 \pm 1$ & $611 \pm 1$ &  & 58349.0 & d & $-502 \pm 2$ & $-458 \pm 1$ &  \\
59123.5 & b & $1620 \pm 1$ & $591 \pm 3$ &  & 58786.9 & d & $-527 \pm 3$ & $-432 \pm 3$ &  \\
59448.6 & b & $1626 \pm 1$ & $578 \pm 2$ &  & 58786.9 & d & $-532 \pm 3$ & $-425 \pm 2$ &  \\
51117.0 & c & $-837 \pm 26$ & $483 \pm 23$ &  & 58790.6 & d & $-528 \pm 1$ & $-435 \pm 2$ &  \\
53199.0 & c & $-739 \pm 6$ & $612 \pm 6$ &  & 58790.6 & d & $-530 \pm 2$ & $-430 \pm 1$ &  \\
53568.1 & c & $-713 \pm 5$ & $630 \pm 5$ &  & 59123.5 & d & $-551 \pm 1$ & $-415 \pm 4$ &  \\
54312.7 & c & $-683 \pm 4$ & $671 \pm 4$ &  & 59448.6 & d & $-569 \pm 1$ & $-390 \pm 2$ &  \\
54396.6 & c & $-678 \pm 7$ & $676 \pm 7$ &  & 59448.6 & d & $-563 \pm 2$ & $-391 \pm 2$ &  \\
54656.3 & c & $-658 \pm 4$ & $701 \pm 4$ &  & 55043.7 & e & $-306 \pm 7$ & $-211 \pm 7$ &  \\
54689.3 & c & $-657 \pm 2$ & $706 \pm 2$ &  & 55113.0 & e & $-306 \pm 7$ & $-217 \pm 7$ &  \\
54725.9 & c & $-663 \pm 3$ & $693 \pm 3$ &  & 55134.9 & e & $-310 \pm 9$ & $-187 \pm 9$ &  \\
54791.7 & c & $-654 \pm 20$ & $700 \pm 20$ &  & 55138.6 & e & $-304 \pm 10$ & $-196 \pm 10$ &  \\
54839.3 & c & $-612 \pm 30$ & $665 \pm 30$ &  & 55390.4 & e & $-323 \pm 6$ & $-166 \pm 6$ &  \\
55043.7 & c & $-639 \pm 4$ & $712 \pm 4$ &  & 55397.7 & e & $-329 \pm 6$ & $-178 \pm 6$ &  \\
55087.5 & c & $-634 \pm 30$ & $697 \pm 30$ &  & 55499.9 & e & $-341 \pm 16$ & $-143 \pm 16$ &  \\
55109.4 & c & $-636 \pm 40$ & $692 \pm 40$ &  & 55762.7 & e & $-352 \pm 8$ & $-130 \pm 8$ &  \\
55113.0 & c & $-627 \pm 7$ & $716 \pm 7$ &  & 55850.3 & e & $-326 \pm 13$ & $-119 \pm 13$ &  \\
55134.9 & c & $-636 \pm 9$ & $720 \pm 9$ &  & 55875.9 & e & $-382 \pm 16$ & $-127 \pm 16$ &  \\
55138.6 & c & $-630 \pm 13$ & $720 \pm 13$ &  & 56128.3 & e & $-373 \pm 8$ & $-84 \pm 8$ &  \\
55390.4 & c & $-619 \pm 4$ & $728 \pm 4$ &  & 56227.1 & e & $-370 \pm 9$ & $-76 \pm 9$ &  \\
55397.7 & c & $-606 \pm 6$ & $725 \pm 6$ &  & 56230.8 & e & $-371 \pm 6$ & $-80 \pm 6$ &  \\
55499.9 & c & $-607 \pm 12$ & $744 \pm 12$ &  & 56581.3 & e & $-373 \pm 13$ & $-17 \pm 13$ &  \\
55762.7 & c & $-595 \pm 4$ & $747 \pm 4$ &  & 56588.6 & e & $-394 \pm 11$ & $-36 \pm 17$ &  \\
55850.3 & c & $-561 \pm 13$ & $752 \pm 13$ &  & 56614.2 & e & $-381.2 \pm 2.2$ & $-30.3 \pm 1.5$ & \commenta \\
55875.9 & c & $-578 \pm 13$ & $767 \pm 13$ &  & 56851.4 & e & $-389 \pm 1$ & $-22 \pm 2$ &  \\
56128.3 & c & $-578 \pm 5$ & $761 \pm 5$ &  & 56851.4 & e & $-387 \pm 2$ & $-10 \pm 3$ &  \\
56227.1 & c & $-572 \pm 3$ & $768 \pm 3$ &  & 56855.1 & e & $-387 \pm 11$ & $3 \pm 11$ &  \\
56230.8 & c & $-557 \pm 6$ & $763 \pm 6$ &  & 56884.3 & e & $-389 \pm 1$ & $-17 \pm 2$ &  \\
56581.3 & c & $-542 \pm 22$ & $784 \pm 22$ &  & 56997.5 & e & $-385 \pm 3$ & $12 \pm 2$ &  \\
56588.6 & c & $-538 \pm 6$ & $784 \pm 13$ &  & 56997.5 & e & $-389 \pm 8$ & $11 \pm 4$ &  \\
56614.2 & c & $-537.4 \pm 1.8$ & $782.3 \pm 1.6$ & \commenta & 56997.5 & e & $-390 \pm 7$ & $12 \pm 4$ &  \\
56851.4 & c & $-521 \pm 3$ & $790 \pm 9$ &  & 56997.5 & e & $-383 \pm 8$ & $11 \pm 4$ &  \\
56855.1 & c & $-540 \pm 13$ & $799 \pm 13$ &  & 57209.1 & e & $-391 \pm 1$ & $33 \pm 2$ &  \\
56913.5 & c & $-519.3 \pm 2.3$ & $793.8 \pm 2.0$ & \commenta & 57209.1 & e & $-383 \pm 9$ & $33 \pm 5$ &  \\
56997.5 & c & $-511 \pm 2$ & $799 \pm 2$ &  & 57234.7 & e & $-386 \pm 1$ & $36 \pm 1$ &  \\
56997.5 & c & $-514 \pm 3$ & $798 \pm 4$ &  & 57293.1 & e & $-392 \pm 1$ & $39 \pm 3$ &  \\
56997.5 & c & $-512 \pm 3$ & $798 \pm 4$ &  & 57293.1 & e & $-382 \pm 9$ & $50 \pm 5$ &  \\
56997.5 & c & $-512 \pm 3$ & $797 \pm 4$ &  & 57651.5 & e & $-376.1 \pm 1.8$ & $80.6 \pm 1.7$ & \commenta \\
57209.1 & c & $-498 \pm 1$ & $806 \pm 1$ &  & 57710.1 & e & $-382 \pm 2$ & $94 \pm 6$ &  \\
57234.7 & c & $-495 \pm 2$ & $806 \pm 2$ &  & 57710.1 & e & $-378 \pm 4$ & $90 \pm 2$ &  \\
57260.2 & c & $-482 \pm 5$ & $813 \pm 6$ &  & 57918.3 & e & $-377 \pm 1$ & $115 \pm 3$ &  \\
57293.1 & c & $-494 \pm 1$ & $811 \pm 1$ &  & 57918.3 & e & $-373 \pm 3$ & $118 \pm 2$ &  \\
57651.5 & c & $-465.7 \pm 2.4$ & $820.9 \pm 2.0$ & \commenta & 58038.7 & e & $-372 \pm 1$ & $129 \pm 2$ &  \\
57710.1 & c & $-464 \pm 2$ & $824 \pm 2$ &  & 58038.7 & e & $-369 \pm 1$ & $128 \pm 1$ &  \\
57918.3 & c & $-449 \pm 1$ & $835 \pm 1$ &  & 58038.7 & e & $-370 \pm 2$ & $135 \pm 3$ &  \\
58038.7 & c & $-441 \pm 1$ & $839 \pm 1$ &  & 58038.7 & e & $-371 \pm 2$ & $128 \pm 2$ &  \\
58038.7 & c & $-441 \pm 1$ & $839 \pm 1$ &  & 58286.9 & e & $-360 \pm 2$ & $162 \pm 2$ &  \\
58286.9 & c & $-424 \pm 1$ & $848 \pm 1$ &  & 58286.9 & e & $-358 \pm 2$ & $156 \pm 2$ &  \\
58349.0 & c & $-421 \pm 1$ & $850 \pm 1$ &  & 58349.0 & e & $-361 \pm 2$ & $166 \pm 2$ &  \\
58349.0 & c & $-421 \pm 1$ & $851 \pm 1$ &  & 58349.0 & e & $-357 \pm 1$ & $162 \pm 2$ &  \\
58786.9 & c & $-392 \pm 2$ & $875 \pm 2$ &  & 58349.0 & e & $-359 \pm 1$ & $167 \pm 2$ &  \\
58790.6 & c & $-388 \pm 2$ & $870 \pm 2$ &  & 58349.0 & e & $-358 \pm 2$ & $163 \pm 1$ &  \\
59123.5 & c & $-364 \pm 2$ & $883 \pm 1$ &  & 58359.9 & e & $-357.64 \pm 0.07$ & $163.34 \pm 0.18$ & \commentb \\
59448.6 & c & $-339 \pm 2$ & $890 \pm 2$ &  & 58786.9 & e & $-337 \pm 3$ & $210 \pm 3$ &  \\
51117.0 & d & $133 \pm 35$ & $-533 \pm 34$ &  & 58786.9 & e & $-338 \pm 2$ & $215 \pm 2$ &  \\
53568.1 & d & $-87 \pm 10$ & $-578 \pm 10$ &  & 58790.6 & e & $-337 \pm 2$ & $208 \pm 2$ &  \\
54312.7 & d & $-179 \pm 5$ & $-588 \pm 5$ &  & 58790.6 & e & $-335 \pm 1$ & $210 \pm 1$ &  \\
54396.6 & d & $-175 \pm 10$ & $-589 \pm 10$ &  & 59123.5 & e & $-315 \pm 3$ & $242 \pm 5$ &  \\
54656.3 & d & $-208 \pm 4$ & $-582 \pm 4$ &  & 59448.6 & e & $-292 \pm 2$ & $276 \pm 2$ &  \\
54689.3 & d & $-216 \pm 2$ & $-582 \pm 2$ &  & 59448.6 & e & $-287 \pm 4$ & $272 \pm 2$ &  \\
54725.9 & d & $-202 \pm 4$ & $-588 \pm 4$ &  &  &  &  &  &  \\
\hline
\end{longtable}

\begin{flushleft}
	\footnotesize
	\textbf{Notes.}  Quoted uncertainties are 1$\sigma$. (a) GPI Data points, adopted from \citet{DeRosaEtal2020}; (b) GRAVITY measurement \citep{GRAVITYCollaborationEtal2019}.
\end{flushleft}

\end{appendix}

\end{document}